%% file: arxiv.tex
\documentclass[onecolumn]{IEEEtran}

\usepackage[ruled]{algorithm2e}
\SetAlFnt{\algofont}
\SetAlCapFnt{\algofont}
\SetAlCapNameFnt{\algofont}
\SetAlCapHSkip{0pt}
\IncMargin{-\parindent}

\usepackage{times,amsfonts,amssymb,amsmath,setspace}
\usepackage{graphicx,url,bm,comment,subfigure,bbm,color}

\newcommand{\tgifeps}[3]{
\begin{figure}[tp]
\centering
\includegraphics[width=#1cm]{#2.eps}
\caption{#3}
\label{fig:#2}
\end{figure}
}

\newtheorem{teorema}{Theorem}[section]

\newtheorem{proposizione}{Proposition}

\newcommand{\TC}[1]{T_C #1}
\newcommand{\lambdahat}{\overline{\lambda}}
\newcommand{\diff}{{\rm\,d}}

\newcommand{\Uno}{\mathbb{I}}

\newcommand{\equaref}[1]{(\ref{eq:#1})}

\newcommand{\ba}{\begin{array}}
\newcommand{\ea}{\end{array}}
\newcommand{\be}{\begin{equation}}
\newcommand{\ee}{\end{equation}}

\begin{document}

\title{
A unified approach to the performance \\ analysis of caching systems
\thanks{$^{(*)}$ V. Martina and E. Leonardi are with Dipartimento di Elettronica, Politecnico di Torino, Italy; M. Garetto is with Dipartimento di Informatica, Universit\`{a} di Torino, Italy.}}
\author{
\vspace{-3mm}
Valentina Martina,  
Michele Garetto, 
Emilio Leonardi ${^{(*)}}$ 
}

\maketitle

\begin{abstract}
We propose a unified methodology to analyze the performance
of caches (both isolated and interconnected), by extending and generalizing
a decoupling technique originally known as Che's approximation, which provides
very accurate results at low computational cost. We consider several
caching policies (including very attractive one, called $k$-LRU), 
taking into account the effects of temporal locality.
In the case of interconnected caches, our approach allows us to do better
than the Poisson approximation commonly adopted in prior work.
Our results, validated against simulations and trace-driven experiments,
provide interesting insights into the performance of caching systems. 
\end{abstract}

\input{intro_tech}
\input{preliminaries}

\input{single_cache}

\input{network}

\input{related}

\section{Conclusions}\label{sec:concl}
The main goal of this paper was to show that a variety
of caching systems (both isolated and interconnected caches)
operating under various insertion/eviction policies
and traffic conditions, can be accurately analysed within
a unified framework based on a fairly general decoupling 
principle extending the original Che's approximation.
We have also shown that many properties of cache systems
can be obtained within our framework 
in a simple and elegant way, including asymptotic results which would otherwise require 
significant efforts to be established.
From the point of view of system design, our study 
has revealed the superiority of the k-LRU policy, in terms
of both simplicity and performance gains.
Still many extensions and refinements are possible, 
especially for cache networks under general traffic.


\appendix
\section*{APPENDIX}
\input{appendix}

\input{app_2LRUp_in}

\bibliographystyle{ieeetr}
\bibliography{arxiv}


\end{document}

%% file: intro_tech.tex
\section{Introduction and paper contributions}
In the past few years the performance of caching systems, one of the
most traditional and widely investigated topic in computer science,
has received a renewed interest by the networking research community.
This revival can be essentially attributed to the crucial role
played by caching in new content distribution systems emerging in the 
Internet. 
Thanks to an  impressive proliferation of proxy servers, Content Delivery Networks (CDN) 
represent today the standard solution adopted by content providers to serve large 
populations of geographically spread users~\cite{Jiang_conext12}.
By caching contents close to the users, we jointly reduce
network traffic and improve user-perceived experience.

The fundamental role played by caching systems in the Internet 
goes beyond existing content delivery networks,
as consequence of the gradual shift from the traditional host-to-host
communication model to the new host-to-content paradigm. Indeed, a novel Information-Centric 
Networking (ICN) architecture has been proposed for the future Internet 
to better respond to the today and future (according to  predictions) traffic 
characteristics~\cite{Jacobson-ICN}. In this architecture, 
caching becomes an ubiquitous functionality available at each router.

For these reasons it is of paramount importance to develop efficient tools
for the performance analysis of large-scale interconnected caches 
for content distribution.
Unfortunately, evaluating the performance of cache networks is hard,
considering that the computational cost to exactly analyse just a single LRU 
(Least Recently Used) cache, grows exponentially with both the cache size and the number of contents
\cite{King,Dan1990}. Nevertheless, several approximations have been
proposed over the years \cite{Dan1990,Che,Roberts1,kurose2010,muscariellosig,amiciburini}
which can accurately predict cache performance at 
an affordable computational cost.

The main drawback of existing analytical techniques is their rather limited scope. 
Indeed, many of them target only specific caching policies (mainly LRU and FIFO)
under simplifying traffic conditions (most of previous work relies on 
the Independent Reference Model \cite{Coffman:73}), while the analysis of cache networks
has only recently been attempted (essentially for LRU) -- see related work
in Sec. \ref{sec:related}.

The main contribution of our work is to show that the decoupling principle
underlying one of the approximations suggested in the past (the so called 
Che approximation) has much broader applicability than the particular context 
in which it was originally proposed (\emph{i.e.}, a single LRU cache under IRM traffic),
and can actually provide the key to develop a general
methodology to analyse a variety of caching systems.

{
In particular, in this paper we show how to extend and generalize  
the decoupling principle of Che's approximation along three orthogonal directions: i) 
a much larger set of caching algorithms than those analysed so far (under Che's approximation),
implementing different insertion/eviction policies (including a multi-stage LRU scheme,  
LRU with probabilistic insertion, FIFO and RANDOM); 
ii) more general traffic model than the traditional
IRM, so as to capture the effects of temporal locality in the 
requests arrival process (in particular, we consider a general renewal traffic model
for all the above-mentioned caching policies); iii) a more accurate technique to analyse 
interconnected caches that goes beyond the standard Poisson assumption 
adopted so far, and permits considering also smart replication 
strategies (such as leave-copy-probabilistically and leave-copy-down).}

Although in this paper we cannot analyse all possible combinations
of the above extensions, we provide sufficient evidence that a unified
framework for the performance analysis of caching systems
is indeed possible under the Che approximation at low
computational cost. Our results for the considered systems 
turn out to be surprisingly good when compared to simulations (model predictions
can be hardly distinguished from simulation results on almost all plots). 
 
 
Furthermore, under the small cache regime (\emph{i.e.}, cache size small with 
respect to the content catalogue size), which is of special interest for ICN,
our expressions can be further simplified, leading to simple closed-form 
formulas for the cache hit probability, revealing 
interesting asymptotic properties of the various caching policies.
The insights gained from our models are also (qualitatively) confirmed
by trace-driven experiments.
 
To the best of our knowledge, we are the first to propose 
a unified, simple and flexible approach that can be used as the basis 
of a general performance evaluation tool for caching systems.

{This paper extends the previous conference version under several 
respects: i) our modeling approach has been generalized and successfully applied to cache networks with general (mesh) topology;  ii) 
new material concerning the asymptotic behavior of some of the considered caching policies 
has been added; iii) several parts of have been modified to improve the overall clarity.}

%% file: preliminaries.tex
\section{System assumptions}\label{sec:prelim}

\subsection{Traffic model}\label{subsec:traffic}
We first recall the so-called Independent Reference Model (IRM), which
is de-facto the standard approach adopted in the literature to 
characterize the pattern of object requests arriving at a cache \cite{Coffman:73}. 
The IRM is based on the following fundamental assumptions: i) users request items
from a fixed catalogue of $M$ object; ii) the probability $p_m$ that a request is for object 
$m$, \mbox{$1 \leq m \leq M$}, is constant (\emph{i.e.}, the object popularity does not vary over time)
and {\em independent} of all past requests, generating an i.i.d. sequence of requests.

{{
By construction, the IRM completely ignores all temporal correlations
in the sequence of requests. In particular, it does not take into
account an important feature often observed in  real content request traces, and typically referred to as {\em
  temporal locality}:  requests for a given content
 become denser over short periods of time. 
The important role played by temporal locality, especially
its beneficial effect on cache performance, is well known in the
context of computer memory architecture~\cite{Coffman:73} and web
traffic~\cite{Fonseca:03}. 
Several extensions of IRM have been already proposed to reproduce content temporal 
locality~\cite{Coffman:73,Fonseca:03,Crovella,Bestavros,nostro-ccr,nostro-infocom15}.  
The majority of the proposed approaches~\cite{Coffman:73,Fonseca:03,Crovella,Bestavros,nostro-infocom15} 
share with the IRM the following two assumptions: i) the content catalog is
fixed; ii) the request process for each content is stationary (typically  it is assumed to be either a renewal process or a semi-Markov-modulated Poisson
process).
Recently~\cite{nostro-ccr} a new traffic model, named Shot Noise Model (SNM),
has been proposed as a viable alternative to traditional traffic models to capture macroscopic
effects related to content popularity dynamics. The  basic idea of the SNM is  to represent the overall request process as the superposition of a potentially infinite population of   independent
inhomogeneous Poisson  processes (shots), each referring to an individual content.
The definition of analytical models for the evaluation of cache performance   under the SNM~\cite{nostro-ccr,olmos}, however,  
is significantly challenging,  as discussed in~\cite{nostro-infocom15}, especially when   non-LRU caches and networks of caches 
are analyzed. Moreover,  in~\cite{nostro-infocom15} it has been shown that the performance of caching system 
under the SNM traffic model can predicted with high accuracy by adopting a fixed-size content catalogue, 
and modeling the arrival process of each content by a renewal process with a specific inter-request time distribution.}

For the above reasons  in this paper we will consider
the following traffic model which generalizes the classical IRM.
The request process for every content $m$ is described by an independent renewal process
with assigned inter-request time distribution. Let $F_R(m,t)$ be the cdf of the inter-request 
time $t$ for object $m$. The average request rate $\lambda_m$ for content $m$
is then given by $\lambda_m = 1/\int_0^\infty(1-F_R(m,t)) \diff t$.
Let $\Lambda = \sum_{m = 1}^{M} \lambda_m$ be the global arrival rate of
requests. 
Note that, by adopting an object popularity law analogous to the one considered 
by the IRM, we also have $\lambda_m = \Lambda p_m$.
 
As a particular case, our traffic model reduces to the classical IRM when 
inter-arrival request times are independently, exponentially distributed, so that requests for object $m$
are generated according to a homogeneous Poisson process of rate $\lambda_m$.
In the following, we will refer to our generalized traffic model as {\em renewal} traffic.

\subsection{Popularity law}  
Traffic models like the IRM (and its generalizations) are commonly used 
in combination with a Zipf-like law of object popularity,  which is
frequently observed in traffic measurements and widely adopted in
performance evaluation studies~\cite{breslau,zipf}.

In its simplest form, Zipf's law states that the probability to request the $i$-th
most popular item is proportional to $1/i^\alpha$, where the exponent
$\alpha$ depends on the considered system (especially on the type of
objects), and plays a crucial role on the resulting 
cache performance~\cite{Roberts1}. Estimates of $\alpha$ reported in the 
literature for various kinds of systems range between .65 and 1 ~\cite{Roberts_mix}.

In our work, we will consider a simple Zipf's law as the
object popularity law, although our results hold
in general, \emph{i.e.}, for any given distribution of
object request probabilities $\{p_m\}_m$.

\subsection{Policies for individual caches}
There exists a tremendous number of different policies
to manage a single cache, which differ either for the insertion or for 
the eviction rule. 
We will consider the following algorithms, as a representative
set of existing policies:
\begin{itemize}
\item {\bf LFU}: the Least Frequently Used policy statically stores in the cache the $C$ most 
popular contents (assuming their popularity is known a-priori); LFU is known to provide 
optimal performance under IRM. 
\item {\bf LRU}: upon arrival of a request, an object not already stored in the cache is inserted into it.
If the cache is full, to make room for a new object the {\em Least Recently Used} item is evicted, \emph{i.e.}, the
object which has not been requested for the longest time.
\item {\bf q-LRU}: it differs from LRU for the insertion policy: upon arrival of a request, 
an object not already stored in the cache is inserted into it with probability $q$. The eviction policy
is the same as LRU.
\item {\bf FIFO}: it differs from LRU for the eviction policy: to make room
for a new object, the item inserted the longest time ago is evicted. 
Notice that this scheme differs from LRU in this respect: requests finding an object in the 
cache do not \lq refresh' the arrival time associated to it.  
\item{\bf RANDOM}:  it differs from LRU for the eviction policy: 
to make room for a new object, a random item stored in the cache is evicted.
\item{\bf k-LRU}: {{this strategy provides a clever insertion policy by exploiting
the following idea: before arriving at the (physical) cache which is storing actual objects,  
indexed by $k$, requests have to advance through a chain of $k-1$  (virtual) caches put in front of it, 
acting as filters, which store only object pointers performing caching operations on them (see Fig. \ref{fig:klru_4}).
Specifically, upon arrival of a request, a content/pointer can be stored in cache $i>1$ only if its pointer
is already stored in cache $i-1$ (i.e. the arrival request has produced a hit in cache $i-1$).
The eviction policy at all caches is LRU.
We remark that this policy\footnote{In the most general case one could individually specify the size of all caches along the chain; however, for simplicity, in this  paper we will
restrict ourselves to the case in which all caches have the same size (expressed either in terms of objects or pointers), 
since numerical explorations suggest that no significant performance gains can be obtained 
by tuning the sizes of individual caches.} can be seen as a generalization of the two-stages policy 
proposed in~\cite{LRU-2Q}}, called there LRU-2Q. 
\item {\bf k-RANDOM}: it works exactly like k-LRU,  
with the only difference that the eviction policy at each cache is \mbox{RANDOM}.}
\end{itemize}

\tgifeps{12}{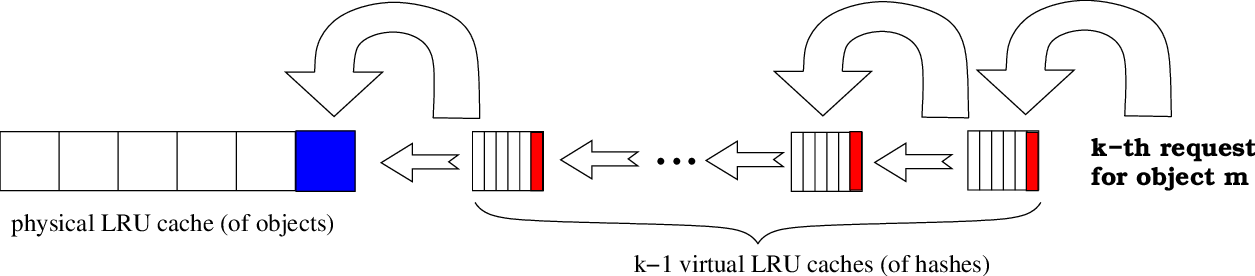}{Illustration of k-LRU policy.}

We remark that LRU has been widely adopted, since it provides good performance while 
being reasonably simple to implement. RANDOM and FIFO have been considered as 
viable alternative to LRU in the context of ICN, as their hardware implementation
in high-speed routers is even simpler. The q-LRU policy and multi-stage 
caching systems similar to our k-LRU have been proposed in the past to 
improve the performance of LRU by means of a better insertion policy.
We have chosen q-LRU in light of its simplicity, and the 
fact that it can be given an immediate interpretation 
in terms of probabilistic replication for cache networks (see next section).
The main strength of k-LRU, instead, resides in the fact that it requires
just one, traffic-independent parameter\footnote{More sophisticated insertion policies 
such as the persistent-access-caching algorithm \cite{jel08}
obtain a filtering effect similar to $k$-LRU but 
require more parameters which are not easy to set, 
requiring a-priori knowledge of the traffic characteristics.} 
(the number of caches $k$), providing significant improvements 
over LRU even for very small $k$ (much of the possible gain is 
already achieved by $k=2$).

\subsection{Replication strategies for cache networks}\label{subsec:rep}
In a system of interconnected caches, requests
producing a miss at one cache are typically forwarded along 
one or more routes toward repositories storing all objects.
After the request eventually hits the target, we need to specify how 
the object gets replicated back in the 
network, in particular along the route traversed by the  
request. We will consider the following mechanisms \cite{rossi-icn14}:
\begin{itemize}
\item {\bf leave-copy-everywhere (LCE)}: the object is 
sent to all caches of the backward path.
\item {\bf leave-copy-probabilistically (LCP)}: the object is sent with 
probability $q$ to each cache of the backward path.
\item {\bf leave-copy-down (LCD)}: the object is sent only to the 
cache preceding the one in which the object is found (unless the object is found
in the first visited cache).
\end{itemize}
Notice that LCP, combined with standard LRU at all caches, is the same as
LCE combined with q-LRU at all caches.

\section{The Che approximation} \label{sect:che-approx}
We briefly recall Che's approximation for LRU  under the classical
IRM \cite{Che}. 
Consider a cache capable of storing $C$ objects. Let
$T_C(m)$ be the time needed before $C$ {\em distinct} objects (not including
$m$) are requested by users. Therefore, $T_C(m)$ is the {\em
cache eviction time} for content $m$, \emph{i.e.}, the time since the last request after 
which object $m$ will be evicted from the cache (if the object is not again requested 
in the meantime).  

Che's approximation assumes $T_C(m)$ to be
a constant independent of the selected content $m$.
This assumption has been given a theoretical justification  recently 
in~\cite{Roberts1}, where it is
shown that, under a Zipf-like popularity distribution, the
coefficient of variation of the random variable representing $T_C(m)$
tends to vanish as the cache size grows. Furthermore, the dependence
of the eviction time on $m$ becomes negligible when the catalogue size 
is sufficiently large. { For completeness we wish to remark that an indirect proof  of Che's approximation asymptotic validity has been 
provided  earlier in  \cite{Jelenkovic:2008:proof}  for $\alpha>1$.} 

The reason why Che's approximation greatly simplifies the analysis of 
caching systems is because it allows to decouple 
the dynamics of different contents: interaction among the 
contents is summarized by $T_C$, which acts as a single 
primitive quantity representing the response of the cache
to an object request.


More in detail, thanks to Che's approximation, we can state that 
an object $m$ is in the cache at time $t$, if and only if a time
smaller than $T_C$ has elapsed since the last request for object $m$,
\emph{i.e.}, if at least one request for $m$ has arrived in the
interval $(t-T_c,t]$. Under the assumption that requests for object $m$ arrive according 
to a Poisson process of rate $\lambda_m$, 
the time-average probability $p_{\text{in}}(m)$ that object $m$ 
is in the cache is then given by:  
\begin{equation}\label{eq:phm}
p_{\text{in}}(m) = 1- e^{-\lambda_{m} T_c} 
\end{equation}
As immediate consequence of PASTA property for Poisson arrivals, 
observe that $p_{\text{in}}(m)$ represents, by construction, 
also the hit probability $p_{\text{hit}}(m)$,  \emph{i.e.}, the probability that a 
request for object $m$ finds object $m$ in the cache.   

Considering a cache of size $C$, by construction: 
\[ 
C= \sum_{m} \Uno_{\{\text{$m$ in cache}\}} 
\]
After averaging both sides, we obtain:
\be\label{eq:C}
C= \sum_{m} \mathbb{E}[\Uno_{\{\text{$m$ in cache}\}} ]= \sum_m p_{\text{in}}(m).
\ee
The only unknown quantity in the above equality is $T_C$,
which can be obtained with arbitrary precision by a fixed point procedure.
The average hit probability of the cache is:
\begin{equation}\label{eq:phmtot}
p_{\text{hit}}= \sum_{m} p_m \, p_{\text{hit}}(m)
\end{equation}



%% file: single_cache.tex
\section{Extensions for single cache}\label{sec:single}
We will show in the next sections that Che's idea of summarizing the interaction among 
different contents by a single variable (the cache eviction time) 
provides a powerful decoupling technique that can be used 
to predict cache performance also under {\em renewal} traffic, as well as to  
analyze policies other than LRU. 

\subsection{LRU under {\em renewal} traffic}
The extension of Che's approximation to the {\em renewal} traffic model 
is conceptually simple although it requires some care.  
Indeed, observe that, under a  general request process, we can not apply 
PASTA anymore, identifying $p_{\text{in}}(m)$
with  $p_{\text{hit}}(m)$. 
To compute $p_{\text{in}}(m)$  we can still consider that an object
$m$ is in the cache at time $t$ if and only if the last request arrived in $[t-T_C, t)$. 
This requires that the {\em age} since the last request for object $m$ is smaller than $T_C$:
\[
p_{\text{in}}(m)=\hat{F}_R(m,T_C) 
\]
where { $\hat{F}_R(m,t)=\lambda_m\int_0^t (1- F_R(m,\tau)) \diff \tau $ } is the cdf of the {\em age} associated to object-$m$ 
inter-request time distribution.

On the other hand, when  computing  $p_{\text{hit}}(m)$, we implicitly condition on the fact 
that a request arrives at time $t$. Thus, the probability that the previous request occurred in 
$[t-T_C, t)$ equals the probability that the last inter-request time 
does not exceed  $T_C$, yielding:
\[
p_{\text{hit}}(m)=F_R(m,T_C).
\]

\subsection{q-LRU under IRM and {\em renewal} traffic}\label{sec:qlru}
We now analyse the q-LRU policy (LRU with probabilistic insertion), 
considering first the simpler case of IRM traffic.
In this case, $p_{\text{in}}(m)$ and $p_{\text{hit}}(m)$ are equal by PASTA.

To compute $p_{\text{in}}(m)$ we exploit the following reasoning:
an object $m$ is in the cache at time $t$ provided that: 
i) the last request arrived at $\tau\in[t-T_C,t)$ and ii) either 
at $\tau^-$ object $m$ was already in the cache, or its insertion was 
triggered by the request arriving at $\tau$  (with probability $q$).
We obtain:
\be\label{eq:q-LRU_implicit}
p_{\text{hit}}(m)=p_{\text{in}}(m)=(1- e^{-\lambda_m T_C})[p_{\text{in}}(m)+q(1-p_{\text{in}}(m))] 
\ee
Solving the above expression for $p_{\text{in}}(m)$, we get:
\be\label{eq:q-LRU}
p_{\text{hit}}(m)=p_{\text{in}}(m)=\dfrac{q(1- e^{-\lambda_m T_C})}{e^{-\lambda_m T_C}+q(1- e^{-\lambda_m T_C})}
\ee


Under {\em renewal} traffic, $p_{\text{in}}(m)$ and $p_{\text{hit}}(m)$ differ
by the same token considered for LRU.
Repeating the same arguments as before, we get:
\be\label{eq:qphit-LRU_implicit-gen}
p_{\text{hit}}(m)=F(m,T_C)[p_{\text{hit}}(m)+q(1-p_{\text{hit}}(m))] 
\ee
which generalizes \equaref{q-LRU_implicit}.
The {\em age} distribution must be instead used to compute $p_{\text{in}}(m)$:
\be\label{eq:qpin-LRU_implicit-gen}
p_{\text{in}}(m)=\hat{F}(m,T_C)[p_{\text{hit}}(m)+q(1-p_{\text{hit}}(m))] 
\ee

Regarding the q-LRU policy, Che's approximation allows 
to establish the following interesting property as $q \to 0$,
whose proof is reported in Appendix \ref{app:qlru} (IRM case)
and \ref{app:qlru_general} (non-IRM case).
{
\begin{teorema} \label{theo:qlrugen}
The q-LRU policy tends asymptotically to LFU  
as the insertion probability goes to zero both under IRM and under renewal traffic under the following conditions:  
for any $m_1$ and $m_2$ with 
 $\lambda_{m_1}<\lambda_{m_2}$ either  $\lim_{t\to \infty} \frac{1-F(m_1,t)}{1-F(m_2,t)}=\infty$ or a $T$ can be found such that ${1-F(m_1,T)}>0$ and ${1-F(m_2,T)=0}$.
\end{teorema}
{{\bf  Remark:}  Note that the above condition is satisfied whenever $F(m,t)$  has an exponential tail, i.e., 
$F(m,t)\approx e^{-\alpha_m t}$  with parameter $\alpha_m$ monotonically dependent on the average rate $\lambda_m$;  
instead, it is not satisfied whenever distributions  $F(m,t)$ are power-law, i.e., 
$F(m,t)\approx (\alpha_m t) ^{-k}$ }}.

\subsection{RANDOM and FIFO}
The decoupling principle can be easily extended to RANDOM/FIFO caching policies
by reinterpreting  $T_C(m)$ as the (in general random) sojourn time of content $m$ in the 
cache. In the same spirit of the original Che's approximation, 
we assume $T_C(m)=T_C$ to be a primitive {\em random} variable 
(not any more a constant) whose distribution does not depend on $m$.   

Under IRM traffic the dynamics of each content $m$ 
in the cache can be described by an M/G/1/0 queuing model.  
Indeed observe that object $m$, when not in the cache, 
enters it according to a Poisson arrival process, then it stays in the cache 
for a duration equal to $T_C$, after which it is evicted {\em independently}
of the arrival of other requests for content $m$ during the sojourn time.

The expression of $p_{\text{in}}(m)$ and  $p_{\text{hit}}(m)$ can then be immediately obtained
from Erlang-B formula (exploiting PASTA):
\[
p_{\text{hit}}(m)=  p_{\text{in}}(m)=\lambda_m \mathbb{E}[T_C]/(1+\lambda_m \mathbb{E}[T_C])
\]
Notice that we still employ \eqref{eq:C} to compute $\mathbb{E}[T_C]$.  

As immediate consequence of Erlang-B insensitivity property 
to the distribution of service time, we conclude that,
\begin{proposizione}
Under IRM traffic, the performance of RANDOM and FIFO (in terms of hit probability) are
the same.
\end{proposizione}
This result was originally obtained formally by Gelenbe \cite{Gelenbe}
using a totally different approach that does not resort to Che's approximation.

Note that, under FIFO policy, we can assume $T_C$ to be
a constant, in perfect analogy to LRU. Indeed, $T_C$ is still equal to the time
needed to observe the requests for $C$ distinct objects arriving at the cache.
On the other hand, under RANDOM policy, it is natural to approximate the sojourn
time of an object in the cache with an exponential distribution. Indeed, under RANDOM 
an object is evicted with probability $1/C$ upon arrival of each request for an object 
which is not in the cache.

Under {\em renewal} traffic the dynamics of each object under FIFO and RANDOM can be described, 
respectively, by a G/D/1/0 and a G/M/1/0 queuing model. Observe that, under general traffic,
the performance of FIFO and RANDOM are not necessarily the same.

We now show how the RANDOM policy can be analysed, under {\em renewal traffic}, employing basic queuing theory. 
Probability $p_{\text{hit}}$ can be obtained as the
loss probability of the G/M/1/0 queue. 
Simply put, the hit probability $p_{\text{hit}}(m)$ of a given content $m$ 
equals the probability that the content has not been evicted before the arrival of the 
next request for content $m$. Having approximated the sojourn time in the cache by an exponential distribution, 
we can easily compute:
\[
p_{\text{hit}}(m)= \int_0^{\infty} e^{-r/\mathbb{E}[T_C]} \diff F_{R}(r) = M_R(m,-1/\mathbb{E}[T_C])
\]
where $M_R(m,\cdot)$ is the moment generating function of object-$m$ 
inter-request time.

Probability $p_{\text{in}}(m)$ can also be obtained exploiting the fact that the 
dynamics of a G/M/1/0 system are described by a process that  
regenerates at each arrival. On such a process we can perform a 
standard cycle analysis as follows (we drop the dependency of random variables on $m$ to
simplify the notation).
We denote by $T_{\text{cycle}}$ the duration of a cycle (which corresponds to an inter-request interval). 
Observe that, by construction,  the object is surely in the cache at the beginning of  a cycle.
Let   $\tau$ be the  residual time spent by  the object  in the cache,   since a  cycle has started,
and $T_{\text{ON}}$ be  the time spent by  the object in the cache within a cycle.

By definition, $T_{\text{ON}}=\min\{\tau,T_{\text{cycle}}\}$. Thus,  
by standard renewal theory we have $p_{\text{in}}(m)=\mathbb{E}[T_{\text{ON}}]/\mathbb{E}[T_{\text{cycle}}]$. 
Figure \ref{fig:fig_cycle_2} illustrates the two cases that can occur, 
depending on whether the object is evicted or not before the arrival of the next request.
Now, we know that $\mathbb{E}[T_{\text{cycle}}]= 1/\lambda_m$.
For $\mathbb{E}[T_{\text{ON}}]$, we obtain:
\begin{multline}\label{eq:Ton}
\mathbb{E}[T_{\text{ON}}] = \int_0^{\infty} \left(\mathbb{E}[T_{\text{ON}}\cdot \Uno_{\tau\leq r} \mid 
 T_{\text{cycle}}=r] + \mathbb{E}[T_{\text{ON}}\cdot \Uno_{\tau>r}\mid T_{\text{cycle}}=r]\right) \diff F_{R}(r) = \\
= \int_0^{\infty}\left(\int_0^r \dfrac{x}{{\mathbb{E}[T_C]}}e^{-x/{\mathbb{E}[T_C]}} \diff x
+r e^{-r/{\mathbb{E}[T_C]}}\right) \diff F_{R}(r) 
\end{multline}
In the end, we get $ p_{\text{in}}(m)= \lambda_m \,{\mathbb{E}[T_C]} \,(1-M_R(m,-1/{\mathbb{E}[T_C]})) $.
\tgifeps{9}{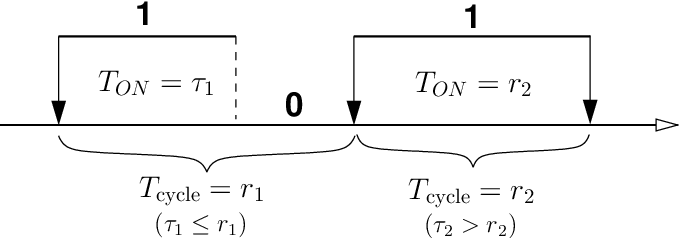}{Illustration of the cycle analysis used for deriving \eqref{eq:Ton}. Vertical arrows represent incoming requests for the content.}
\subsection{2-LRU}
We now move to the  k-LRU strategy, considering first the simple case of $k=2$.
For this system, we derive both a rough approximation based on an additional simplifying assumption
(which is later used to analyse the more general k-LRU) and a more refined model that 
is based only on Che's approximation. For both models we consider
either IRM or {\em renewal} traffic.
 
Let $T^i_C$ be the eviction time of cache $i$. We start observing that meta-cache 1 
behaves exactly like a standard LRU cache, for which we can use
previously derived expressions. 
Under IRM, $p_{\text{in}} (m)$ and $p_{\text{hit}} (m)$ (which are identical by PASTA) 
can be approximately derived by the following argument: object $m$ is found in cache 2 at time $t$ if and only if  
the last request arrived in $\tau\in[t-T^2_C,t)$ and either object $m$ was already in cache 2 at time $\tau^-$ 
or it was not in cache 2 at time $\tau^-$, but its hash was already stored in meta-cache 1. 
Under the additional approximation that the states of meta-cache 1 and cache 2 are independent 
at time $\tau^-$, we obtain:
\begin{equation}\label{eq:rough}
p_{\text{hit}} (m)= p_{\text{in}} (m) \approx 
(1- e^{-\lambda_m T^2_{C})}) \left[ p_{\text{hit}}(m)+(1- e^{-\lambda_m T^1_{C}})
(1-p_{\text{hit}}(m))\right] 
\end{equation}
Observe that the independence assumption between cache 2 and meta-cache 1
is reasonable under the assumption that $T^2_C$ is significantly larger than $T^1_{C}$ (which
is typically the case when the two caches have the same size). Indeed, in this case 
the states of cache 2 and meta-cache 1 tends to de-synchronize, since an hash   
is expunged by meta-cache 1 before the corresponding object is evicted by cache 2, 
making it possible to find an object in cache 2 and not in meta-cache 1
(which otherwise would not be possible if $T^1_{C}\ge T^2_{C}$). 

\tgifeps{5}{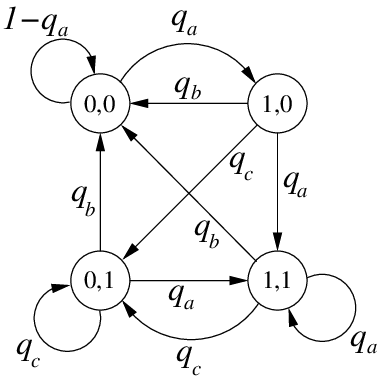}{DTMC describing the dynamics of an object in 2-LRU, sampled at request arrival times }

An exact expression for $p_{\text{hit}} (m)$ (under Che's approximation)  that does not 
require any independence assumption can be derived observing that the dynamics
of object $m$ in the system, sampled at  request arrivals,   can be described by the four states  
Discrete Time Markov Chain (DTMC) represented in Fig. \ref{fig:figMC},
where each state is denoted by a pair of binary variables indicating 
the presence of object $m$ in meta-cache 1 and cache 2, respectively.
Solving the DTMC, we get:
\be \label{eq:2LRU}
p_{\text{hit}} (m)= p_{\text{in}} (m)=1 - \frac{(1+q_a)q_b}{q_a+q_b}
\ee
with $q_a=1-e^{-\lambda_m T_C^1} $, $q_b=e^{-\lambda_m T^2_C}$ and $q_c=1-(q_a+q_b)$.

The extension to {\em renewal} traffic can be carried out following the same lines as before.
Under the additional independence assumption between the two caches, we obtain: 
\begin{eqnarray*}
p_{\text{hit}} (m) &\approx & F_R(m,T^2_C) \left[ p_{\text{hit}}(m)+F_R(m,T^1_C) (1-p_{\text{hit}}(m)) \right] \\
p_{\text{in}} (m) &\approx &\hat{F}_R(m,T^2_C) \left[ p_{\text{hit}}(m)+ F_R(m,T^1_C) (1-p_{\text{hit}}(m))\right] 
\end{eqnarray*}

Also the refined model can be generalized to {\em renewal} traffic, observing that 
object-$m$ dynamics in the system,  sampled at request arrivals ({i.e., logically
 just before a request arrival}), are  still described by a Markov Chain 
with exactly the same structure as in Fig. \ref{fig:figMC}
(only the expressions of transition probabilities change in an obvious way).
Thus we obtain: 
\[
p_{\text{hit}} (m)=1 - \frac{(1+q_a)q_b}{q_a+q_b}
\]
with $q_a=F(m, T^1_C) $ and $q_b=1-F(m, T^2_C)$

To compute $p_{\text{in}}(m)$ we can resort to a cycle analysis, whose details 
are reported in Appendix \ref{app:2LRUp_in}.

\subsection{k-LRU}
Previous expressions obtained for 2-LRU (under the independence assumption between caches)
can be used to iteratively compute the hit probabilities of all caches in a k-LRU system. 
For example, under IRM, we can use \equaref{rough} to relate the hit probability of object $m$ in cache $i$, 
$p_{\text{hit}} (i,m)$, to the hit probability $p_{\text{hit}} (i-1,m)$ of object $m$ in the 
previous cache, obtaining:
\begin{equation}\label{eq:nLRU}
p_{\text{hit}} (i,m) = p_{\text{in}} (i,m) \approx 
(1- e^{-\lambda_m T^i_{C})}) \left[ p_{\text{hit}}(i,m)+( p_{\text{hit}}(i-1,m))(1-p_{\text{hit}}(i,m)) \right] 
\end{equation}
The generalization to {\em renewal} traffic is straightforward.

At last, for large $k$ we can   state:
\begin{teorema}
According to  \eqref{eq:nLRU}  $k$-LRU tends asymptotically to LFU  
as $k\to \infty$ under IRM and {\em renewal} traffic,  as long as the support of the 
inter-request time distribution is unbounded and for any $m_1$ and $m_2$, with 
 $\lambda_{m_1}<\lambda_{m_2}$, it holds $\lim_{t\to \infty} \frac{1-F(m_1,t)}{1-F(m_2,t)}>1$.
\end{teorema}
The proof is reported in Appendix \ref{app:klru}.

\tgifeps{5}{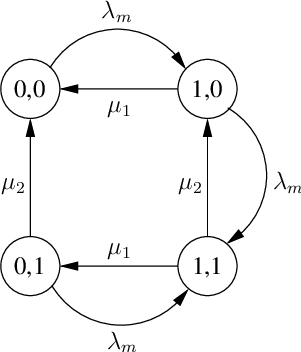}{CTMC describing the dynamics of an object in 2-RANDOM. 
We denoted $\mu_1 = 1/T^1_C$, $\mu_2 = 1/T^2_C$.}

\subsection{k-RANDOM}
Also k-RANDOM can be analysed under Che's approximation assuming exponential sojourn times
in the caches. As an example, the dynamics of an object in 2-RANDOM (under IRM traffic) are described
by the simple four-states continuous time Markov chain depicted in Fig. \ref{fig:fig2random}.
More in general, k-RANDOM can be exactly analyzed by solving a continuous time Markov chain
with $2^k$ states. We omit the details of such standard analysis here. 

\subsection{Small cache approximations}
Small cache approximations can be obtained by replacing the expressions of 
$p_{\text{hit}} (m)$ and $p_{\text{in}} (m)$ with their truncated Taylor expansion
(with respect to $T_C \rightarrow 0$). 
This is especially useful to understand the dependency of $p_{\text{in}}$ and
$p_{\text{hit}}$ on the object arrival rate $\lambda_m$ 
(and thus its popularity), obtaining interesting insights into the 
performance of the various caching policies. 
We restrict ourselves to IRM traffic, however we emphasize that a similar approach 
can be generalized to {\em renewal} traffic. 
We obtain:
\begin{eqnarray*}
\!p_{\text{hit}}(m)\!=\!p_{\text{in}}(m)\!\approx\!
\begin{cases} 
\lambda_m T_C- \frac{(\lambda_m T_C)^2}{2}  & \quad \mbox{LRU }\\
\lambda_m T_C- {(\lambda_m T_C)^2} & \quad \mbox{RANDOM/FIFO }\\
q\lambda_m T_C+ q(\frac{1}{2}-q){(\lambda_m T_C)^2} & \quad \mbox{q-LRU }\\
(\lambda_m)^k \prod_{i=1}^{k} T_C^i & \quad \mbox{k-LRU }
\end{cases}
\end{eqnarray*}
Previous expressions permit us immediately to rank the 
performance of the considered policies in the small 
cache regime. Specifically, better performance is achieved by caching policy
under which $p_{\text{hit}}(m)$ exhibits stronger
dependency on $\lambda_m$. Indeed, recall that (under IRM)
$p_{\text{hit}}=\sum_m \frac{\lambda_m}{\Lambda} p_{\text{hit}}(m)$,
while $\sum_m p_{\text{hit}}(m) = \sum_m p_{\text{in}}(m) = C$.
Hence, the stronger the dependency of $p_{\text{hit}}(m)$ on $\lambda_m$,
the more closely a policy tends to approximate the behavior of LFU (the optimal policy), 
which statically places in the cache the $C$ top popular contents. 

Therefore, k-LRU turns out to be the best strategy, 
since the dependency between $p_{\text{hit}}(m)$  and content popularity 
$\lambda_m$ is polynomial of order $k\ge 2$, in contrast to other policies
(including $q$-LRU for fixed $q$) for which $p_{\text{hit}}(m)$ depends linearly on $\lambda_m$.  
The coefficient of the quadratic term further allows us to rank policies other than k-LRU: 
$q$-LRU is the only policy exhibiting a positive quadratic term (for small $q$), which makes the 
dependency of $p_{\text{hit}}(m)$ on $\lambda_m$ slightly super-linear. 
At last LRU slightly outperforms RANDOM/FIFO because its negative 
quadratic term has a smaller coefficient.


\subsection{Model validation and insights}
The goal of this section is twofold. First, we wish to validate  
previously derived analytical expressions against simulations, 
showing the surprising accuracy of our approximate models 
in all considered cases. Second, we evaluate the impact of system/traffic parameters 
on cache performance, obtaining important insights 
for network design. 

Unless otherwise specified, we will always consider
a catalogue size of $M=10^6$, and a Zipf's law exponent $\alpha=0.8$.

\tgifeps{16.5}{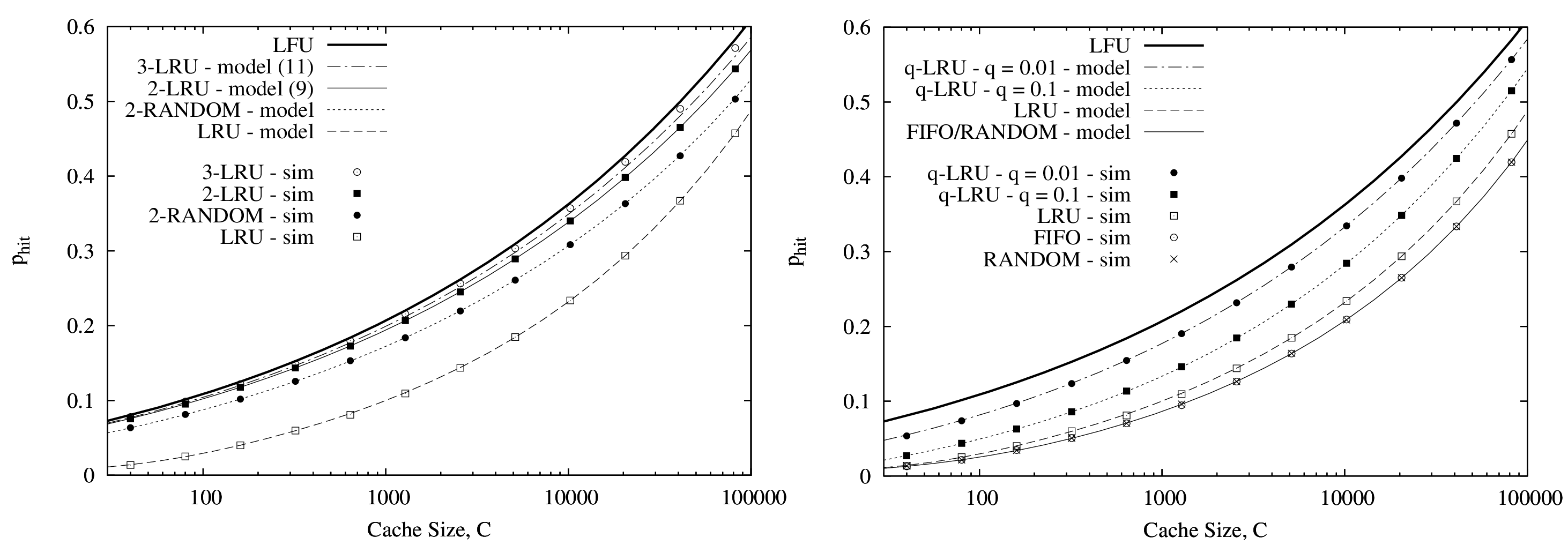}{Hit probability vs cache size, for various
 caching policies, under IRM.}

Fig.~\ref{fig:figura1} reports the hit probability achieved by the 
different caching strategies that we have considered, under IRM traffic. 
Analytical predictions are barely distinguishable from simulation results,
also for the 3-LRU system, for which our approximation \equaref{nLRU}
relies on an additional independence assumption among the caches.

As theoretically predicted, q-LRU (k-LRU) approaches LFU as $q \to 0$ ($k \to \infty$).
Interestingly, the introduction of a single meta-cache in front of an LRU cache (2-LRU)
provides huge benefits, getting very close to optimal performance (LFU).

Differences among the hit probability achieved by the various caching policies
become more significant in the small cache regime (spanning almost 1 order of magnitude). 
In this case, insertion policies providing some protection against unpopular objects
largely outperform policies which do not filter any request.
The impact of the eviction policy, instead, appears to be much weaker, 
with LRU providing moderately better performance than RANDOM/FIFO.  

\tgifeps{9}{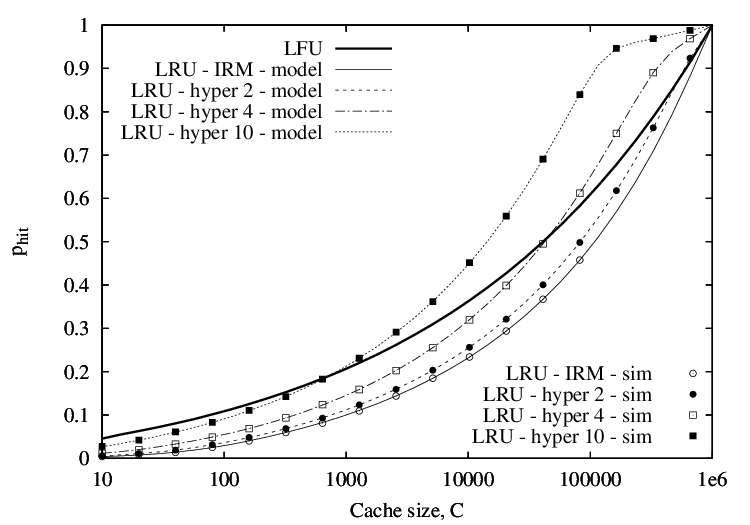}{Hit probability vs cache size, for LRU, under
different degrees of temporal locality.}

Fig.~\ref{fig:figura2} shows the impact of temporal locality on caching performance:   
LRU is evaluated under {\em renewal} traffic
in which object inter-arrival times are distributed according to a 
second order hyper-exponential with branches $\lambda_m^1 = z \lambda_m$ and 
$\lambda_m^2 =  \lambda_m/z$ (hereinafter, we will call hyper-$z$ such distribution),
so that increasing values of $z$ results into stronger temporal locality in 
the request process. We observe that temporal locality can have a dramatic (beneficial) 
impact on hit probability, hence it is crucial to take it into account 
while developing analytical models of cache performance. 

Fig.~\ref{fig:figura2} also shows that LFU is no longer optimal when traffic
does not satisfy the IRM. This because LFU 
statically places in the cache the $C$ most popular objects
(on the basis of the {\em average} request rate of contents),
hence the content of the cache is never adapted to instantaneous traffic conditions,
resulting into suboptimal performance.

\tgifeps{16.5}{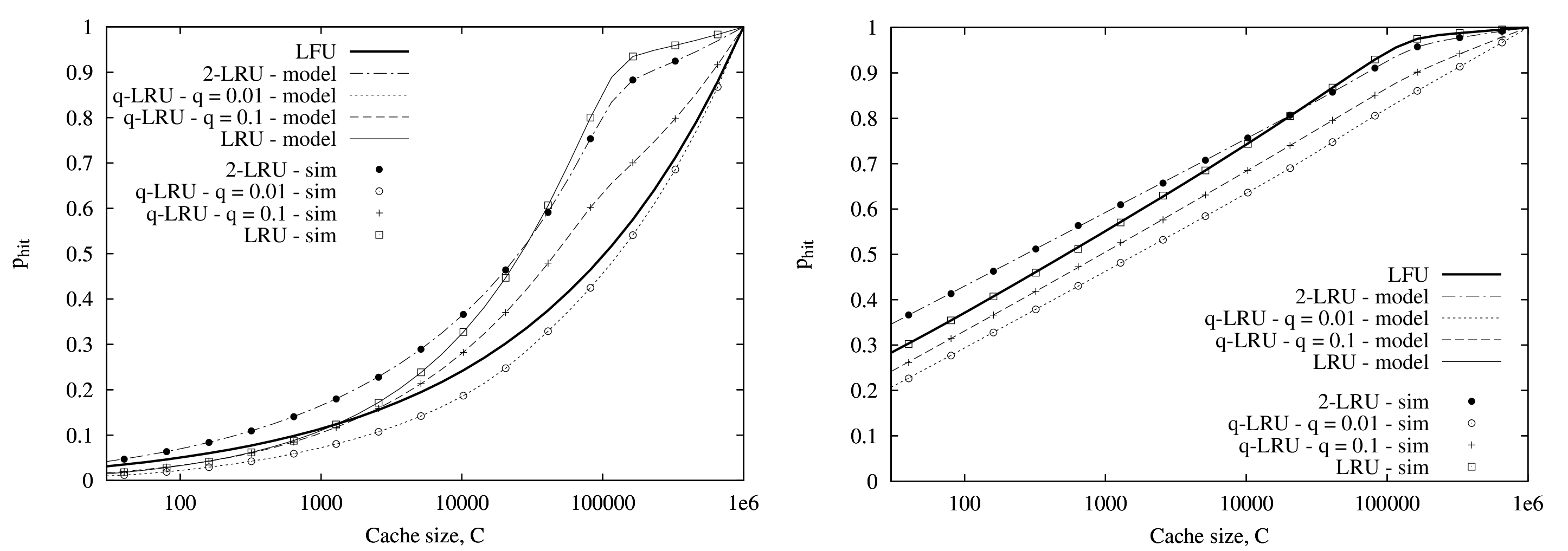}{Hit probability vs cache size, for various
caching policies, under \mbox{hyper-10} traffic, in the case of 
$\alpha=0.7$ (left plot) or $\alpha=1$ (right plot).}

Fig.~\ref{fig:figura3} compares the performance of LFU, LRU, q-LRU and 2-LRU 
in the case in which traffic exhibits significant temporal locality (hyper-10).   
We also change the Zipf's law exponent, considering either
$\alpha=0.7$ (left plot) or $\alpha=1.0$ (right plot).

We observe that $q$-LRU performs poorly in this case, especially for small values of 
$q$ (in sharp contrast to what we have seen under IRM). This because
$q$-LRU with very small $q$ tends to behave like LFU (keeping  
statically in the cache only the objects with the largest {\em average}
arrival rate), which turns out to be suboptimal as it does not
benefit from the temporal locality in the request process.   

On the contrary, a simple 2-LRU system provides very good performance 
also in the presence of strong temporal locality. 
This because, while 2-LRU is able to filter out unpopular contents, its 
insertion policy is fast enough to locally adapt to short-term 
popularity variations induced by temporal locality.

\tgifeps{9}{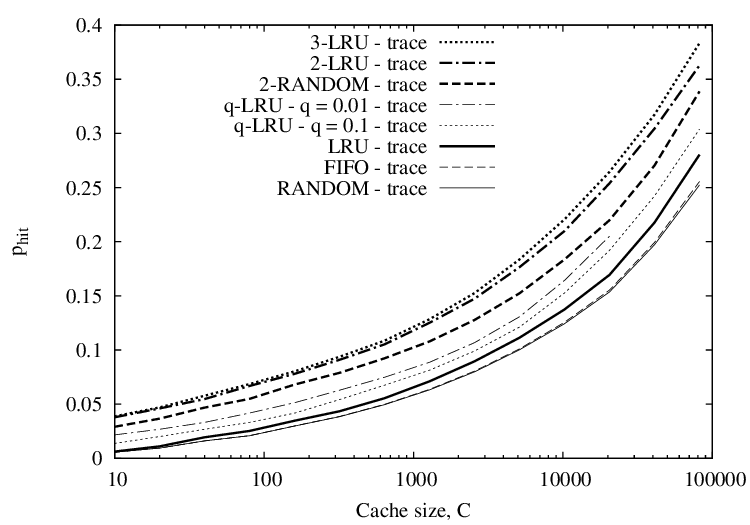}{Hit probability vs cache size, for various
caching policies, under real trace of Youtube video requests.}

To further validate the design insights gained
by our analysis, we have also run a trace-driven experiment, using 
a real trace of Youtube video requests collected
inside the network of a large Italian ISP, offering
Internet access to residential customers.  
The trace has been extracted analysing TCP flows 
by means of Tstat, an open-source traffic monitoring tool developed
at Politecnico di Torino \cite{tstat}. 
During a period of 35 days in year 2012, from March 20th to April 25th, we recorded 
in total 3.8M of requests, for 1.76M of
videos, coming from 31124 distinct IP addresses.

Fig.~\ref{fig:figura6} reports the hit probability achieved by different 
caching schemes\footnote{The largest cache size that we could consider was limited by
the finite duration of the trace.}.    
We observe that most considerations drawn under 
synthetic traffic (in particular, the policy ranking) 
still hold when the cache is fed by real traffic taken from an 
operational network. We summarize the main findings:
i) the insertion policy plays a crucial role in cache performance, especially
in the small-cache regime; 
ii) a single meta-cache (2-LRU system) significantly outperforms
the simple LRU and its probabilistic version (q-LRU), while additional 
meta-caches provide only minor improvements;
iii) the impact of the eviction policy is not significant, 
especially when caches are small with respect to the catalogue size.


%% file: network.tex
\section{Cache networks}\label{sec:network}
In a typical cache network, caches forward their miss stream 
(\emph{i.e.}, requests which have not found the target object) to other caches.
Let us briefly recall the standard approach that has been 
proposed in the literature to analyse this kind of system.

We first introduce some notation.
Let $p_{\text{hit}}(i,m)$ be the hit probability of object $m$
in cache $i$, and $p_{\text{in}}(i,m)$ be the (time average) 
probability that object $m$ is in cache $i$.
We denote by $T^i_C$ the eviction time of cache $i$.
Furthermore, let $\lambdahat_m(i)$ be the total {\em average} arrival rate of
requests for object $m$ at cache $i$. This rate can be immediately
computed, provided that we know the hit probability of object $m$
at all caches sending their miss stream to cache $i$ -- see later equation \equaref{lambdahat}.

Once we know the average arrival rates $\lambdahat_m(i)$, we can simply assume that 
the arrival process of requests for each object 
at any cache is Poisson, and thus independently solve
each cache using its IRM model. A multi-variable 
fixed-point approach is then used to solve the entire system 
(see \cite{kurose2010} for a dissection of the errors introduced by 
this technique).

We now explain how Che's approximation can be exploited
to obtain a more accurate analysis of the cache network,
under the three replication strategies defined in Sec. \ref{subsec:rep}.
To describe our improved technique, it is sufficient to consider
the simple case of just two caches (tandem network). 
Indeed, the extension of our method to general network
is straightforward. 

Moreover, we will limit ourselves to the case of {networks of LRU caches} in which the traffic produced 
by the users satisfies the IRM model (\emph{i.e.}, the exogenous process
of requests for each object is Poisson).
The general idea is to try to capture (though still in an approximate way)
the existing correlation among the states of neighboring caches, which 
is totally neglected under the Poisson approximation.
To do so, a different approximation is needed for each
considered replication strategy, as 
explained in the following sections.

\subsection{Leave-copy-everywhere}\label{subsec:lce}
Focusing on the basic case of a tandem network, the arrival
process of requests for object $m$ at the first cache is
an exogenous Poisson process of rate $\lambda_m(1)$. 
The first cache (which is not influenced by the second one) 
can then be solved using the standard IRM model, giving
\be\label{eq:LRU_1}
p_{\text{hit}}(1,m)=p_{\text{in}}(1,m)=1-e^{-\lambda_m(1) T^1_C}.
\ee

The arrival process of request for object $m$
at the second cache is not Poisson. It is, instead, an 
ON-OFF modulated Poisson process, where the ON state
corresponds to the situation in which object $m$ is not stored
in cache 1, so that requests for this object are forwarded
to cache 2. Instead, no requests for object $m$ can 
arrive at cache 2 when $m$ is present at cache 1 (OFF state).

The standard approximation would be to compute the average
arrival rate $\lambdahat_m(2)= \lambda_m(1) (1- p_{\text{hit}}(1,m))$
and to apply the IRM model also to the second cache.
Can we do better than this? Actually, yes, at least to compute
the hit probability $p_{\text{hit}}(2,m)$, which can, in practice, be 
very different from $p_{\text{in}}(2,m)$ since PASTA does not apply.
 
We observe that a request for $m$ can arrive at time $t$ at cache 2, 
only if object $m$ is not stored in cache 1 at $t^-$. This implies that 
no exogenous requests can have arrived in the interval $[t- T^1_C, t]$
(otherwise $m$ would be present in cache 1 at time t), hence, a fortiori,
no requests for $m$ can have arrived at cache 2 in the same interval.
 
Now, provided that $T^2_C > T^1_C$, object $m$ is found in cache 2 at time $t$, 
if and only if at least one request arrived at cache 2 within  
the interval $[t-T^2_C, t-T^1_C]$.
During this interval, the arrival process at cache 2 is not Poisson
(it depends on the unknown state of cache 1), and we resort to approximating
it by a Poisson process with rate $\lambdahat_m(2)$, obtaining:
\begin{equation}
\label{eq:improved}
p_{\text{hit}} (2,m) \approx 1- e^{-\lambdahat_m(2) (T^2_C-T^1_C)} 
\end{equation}

Essentially, the improvement with respect to the standard approximation
consists in the term $T^2_C-T^1_C$ in the above equation, in place of $T^2_C$.
If, instead, $T^2_C < T^1_C$, we clearly have $p_{\text{hit}} (2,m)=0$.

Note that the above reasoning cannot be applied to compute 
$p_{\text{in}}(2,m)$ (which is necessary to estimate $T^2_C$),
thus we simply express
\[
p_{\text{in}} (2,m) \approx  1- e^{-\lambdahat_m(2) T^2_C} 
\]
as in the standard IRM model.

\tgifeps{9}{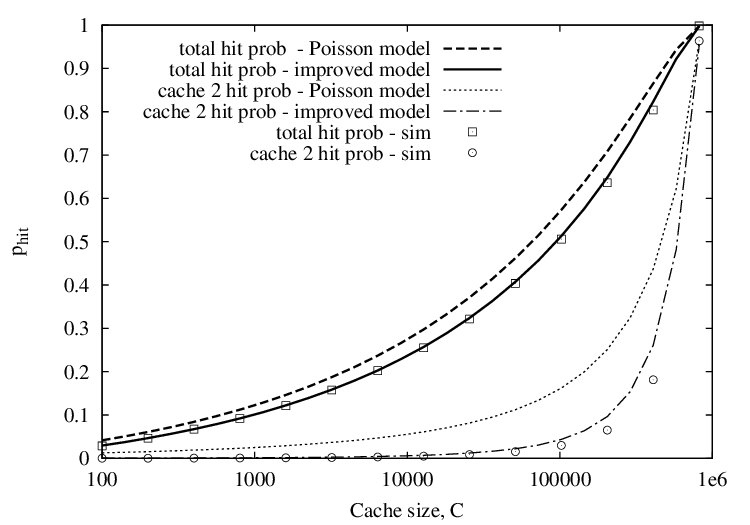}{Comparison between Poisson approximation and our 
improved approximation, in the case of a tandem network of two LRU caches, 
under IRM traffic}

To show the significant gains in terms of accuracy 
that can be obtained by applying our simple improved approximation
with respect to the Poisson approximation, we consider
a tandem network in which the first cache is fed by IRM traffic 
with catalogue size $M=10^6$ and Zipf's law exponent $\alpha=0.8$.
Figure \ref{fig:tandem} reports both the total hit probability
and the hit probability on the second cache, under the 
two considered approximations, against simulation results.
We observe that the Poisson approximation tends to overestimate
the total hit probability, essentially as a consequence 
of a large overestimate of the hit probability on the second cache.  
Our improved approximation, which, recall, essentially 
leads to substituting $T^2_C$ with $T^2_C-T^1_C$
in the standard formula to compute the hit probability of the second cache, 
brings back the analytical prediction 
of total hit probability very close to simulation results,
thanks to a much better model of the behavior of 
the second cache.

\subsection{Leave-copy-probabilistically}\label{subsec:lcp}
Also in this case the first cache is not influenced by the second, hence
we can use the IRM formula of q-LRU \equaref{q-LRU} to analyze its behavior.

To evaluate $p_{\text{hit}} (2,m)$, we observe that
a request for content $m$ that arrives at time $t$ at cache 2  
produces a hit if, and only if, at time $t^-$ content $m$ is stored at cache 2 but not in cache 1.
For this to happen, in the case $T^2_C > T^1_C$ there are two sufficient and necessary 
conditions related to the {\em previous} request for $m$ arriving at cache 2:
i) this request produced a hit at cache 2, or it triggered an insertion here;
ii) it arrived at cache 2 either in the interval  $[t-T^2_C, t-T^1_C]$, or in 
the interval  $[t-T^1_C, t]$  without triggering an insertion in cache 1.
We remark that, in contrast to the LCE case, now it is possible
that the previous request arrived in the interval $[t-T^1_C, t]$: indeed,
the previous request can arrive in this interval, produce a miss in cache 1 (and thus
be forwarded to cache 2) and {\em not} trigger an insertion in cache 1, so that we can
really observe another request arriving at cache 2 at time $t$.
To evaluate the probability of this event, we model  the stream of requests arriving at cache 2 (i.e. producing a miss at cache 1) 
without  triggering an insertion in cache 1 as a Poisson process
with intensity $\lambdahat_m(2) \cdot (1-q)$.
We obtain:
\begin{eqnarray*}
 p_{\text{hit}} (2,m) \approx  [p_{\text{hit}} (2,m) + q(1-p_{\text{hit}} (2,m)) ] \cdot 
\left( 1 - e^{-\lambdahat_m(2) (T^2_C-T^1_C)} \cdot e^{-\lambdahat_m(2) (1-q) T^1_C} \right).
\end{eqnarray*}      
In the above expression, the first term of the product refers to condition i), whereas
the second term account for condition ii) going through the complementary event that
no requests arrive at cache 2 either in the interval $[t-T^1_C, t]$ nor in 
the interval $[t-T^2_C, t-T^1_C]$. Note that this expression reduces to  
\equaref{improved} when $q = 1$ (i.e., LCE).

If, instead, $T^2_C < T^1_C$, the formula simplifies to 
\begin{eqnarray*}
p_{\text{hit}}(2,m) \approx  \left[p_{\text{hit}} (2,m) + q(1-p_{\text{hit}} (2,m))\right] 
\left( 1 - e^{-\lambdahat_m(2) (1-q) T^2_C} \right).
\end{eqnarray*}


To estimate $p_{\text{in}}(2,m)$, we resort to the standard Poisson approximation:
\[
p_{\text{in}}(2,m) \approx (1- e^{-\lambdahat_m(2)T^2_C})\left[p_{\text{in}}(2,m)+q(1-p_{\text{in}}(2, m))\right].  
\] 

\subsection{Leave-copy-down}\label{sec:lcd}
This strategy is more complex to analyse, since now 
the dynamics of cache 1 and cache 2 depend mutually on each other. 
Indeed, it is possible to insert a content in cache 1 only when it is already stored in cache 2.
Probability $p_{\text{in}} (1,m)$ can be computed considering that
object $m$ is found in cache 1 if, and only if, the last request 
arrived in $[t-T^1_C, t]$ and either i) it hit the object in cache 1 or  
ii) it found the object in cache 2 (and not in cache 1). 
Since PASTA holds, we have:
\begin{eqnarray*}
p_{\text{in}} (1,m) \approx p_{\text{hit}}(1,m) = \left[ (1- p_{\text{in}} (1,m)) \,p_{\text{hit}}(2,m)+ p_{\text{in}}(1,m) \right] \cdot ( 1- e^{-\lambda_m(1) \TC(1)})
\end{eqnarray*}
Observe in the previous expression that we have assumed the states of cache 1 and cache 2 to be independent; 
on the other hand, similarly to what we have done before, we write:
\[
p_{\text{in}} (2,m) \approx (1-e^{-\lambdahat_m(2) T^2_C})
\]
Note that, since $p_{\text{in}} (1,m)$ and $p _{\text{in}}(2,m)$ are interdependent, 
a fixed-point iterative procedure is needed to jointly determine them.

It remains to approximate the hit probability at cache 2. 
When $T^2_C > T^1_C$, we write:
\[ 
p_{\text{hit}}(2,m)  \approx   (1- e^{-\lambdahat_m(2) (T^2_C-T^1_C)}) {\,e^{-\lambdahat_m(2) T^1_C}} 
+ (1-e^{-\lambdahat_m(2) \textcolor{black}{(1- p_{\text{hit}}(2,m))} T^1_C})  
\]
Indeed, since at time $t^-$ cache 1 does not store the object by construction, 
either the previous request arrived in $[t-T^2_C, t-T^1_C]$ at cache 2, 
or it arrived  in $[t-T^1_C, t]$ { (again at cache 2)} but it did not trigger an insertion in cache 1 
because object $m$ was not found in cache 2.}
As before, we model  the stream of requests arriving at cache 2 (i.e. producing a miss at cache 1) 
without  triggerring an insertion in the first cache  as a Poisson process
with intensity $\lambdahat_m(2) \cdot  \textcolor{black}{(1- p_{\text{hit}}(2,m))}$.

Similarly, if $T^2_C < T^1_C$:
%
\[
p_{\text{hit}}(2,m) \approx  (1-e^{-\lambda_m(2)\textcolor{black}{(1- p_{\text{hit}}(2,m))} T^2_C}) 
\]


\subsection{Extension to general cache networks}
Our approach, which has been described above for the simple case of a tandem network, 
can be easily generalized to any network. We limit ourselves to 
explaining how this can be done for the leave-copy-everywhere scheme. 
Let $r_{j,i}$ be the fraction of requests for object $m$ which are
forwarded from cache $j$ to cache $i$ (in the case of a miss in cache $j$).
Observe  that  $[r_{j,i}]$ depends on the routing strategy of requests  
adopted in the network and can be considered 
as a given input to the model.

The average arrival rate of requests for $m$ at $i$ is then 
\begin{equation}\label{eq:lambdahat}
\lambdahat_{m}(i)= \sum_{j} \lambdahat_{m}(j) (1- p_{\text{hit}}(j,m) ) r_{j,i}
\end{equation}
and we can immediately express: 
\[
p_{\text{in}} (i,m) \approx 1 - e^{- \lambdahat_{m}(i) T^i_C} 
\]
resorting to the standard Poisson approximation.

Our refined approach to estimating the hit probability 
can still be applied to the computation of the
conditional probability $p_{\text{hit}} (i,m \mid j)$, which is the
probability that a request for object $m$ hits the object at cache $i$, given that
it has been forwarded by cache $j$. This event occurs if, and only if,
either a request arrived at $i$ from $j$ in the time interval $[t-T^i_C, t-T^j_C]$ 
(provided that $T^i_C > T^j_C$), or at least one request arrived at 
$i$ in the interval $[t-T^i_C, t]$ from another cache (different from $j$).
Thus we write:
\[
p_{\text{hit}} (i,m\mid j) \approx 1- e^{-A_{i,j}} 
\]
where 
$$A_{i,j}= r_{j,i} \lambdahat_m(j)(1- p_{in}(j,m)) \max(0,T^i_C-T^j_C) +
\sum_{k \neq j} r_{k,i}\lambdahat_m(k) (1- p_{in}(k,m)) T^i_C$$
The expression for $p_{\text{hit}} (i,m)$ can then be obtained de-conditioning with respect 
to $j$.

Now, in case of tree-like networks previous expressions can be evaluated 
step-by-step starting from the leaves and going up 
towards the root. In case of general mesh networks,  
a global (multi-variate) fixed-point procedure is necessary.

\subsection{Model validation and insights}
As before, our aim here is to jointly validate our analytical models
against simulation, while getting interesting insights into system behavior.

\tgifeps{16.5}{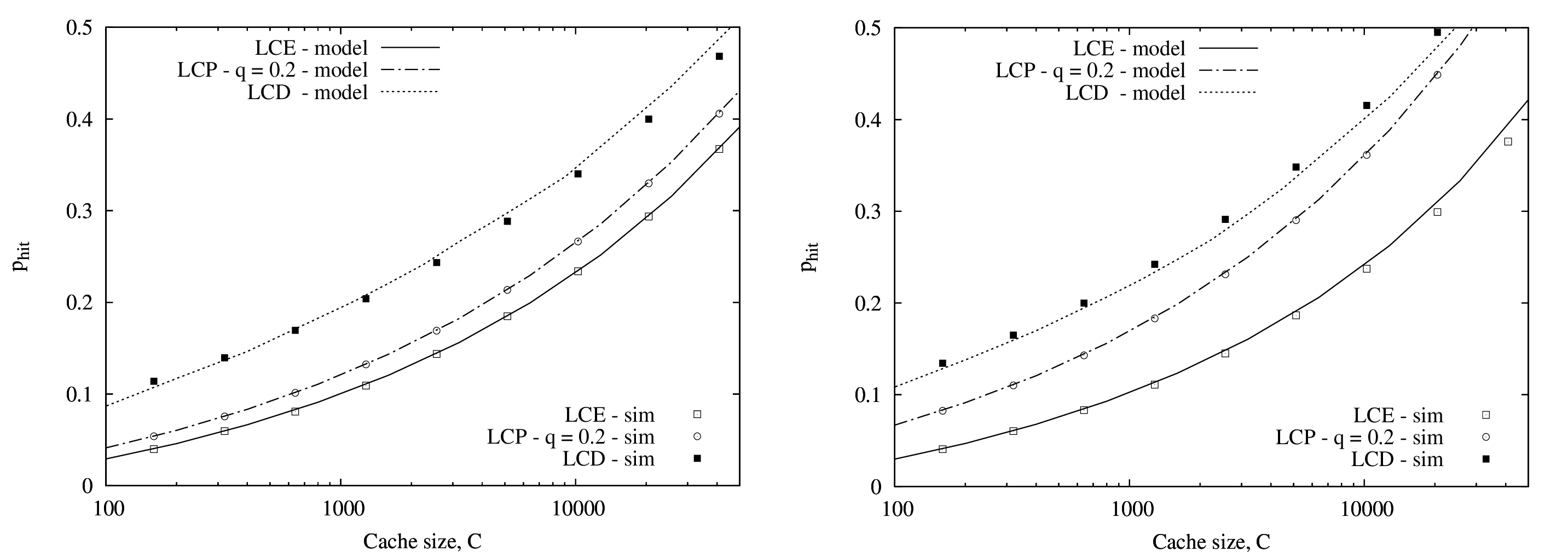}{Hit probability vs cache size, for various
replication strategies, in the case of a chain of 6 caches, under IRM traffic.
Hit probability of the first cache (left plot) and total hit probability of the network
(right plot).}

Fig~\ref{fig:figura4} compares the performance of the different replication strategies
that we have analysed, in the case of a chain of 6 identical caches.
We have chosen a chain topology to validate our model, because
this topology is known to produce the largest degree of correlation
among caches (and thus the maximum deviation from the Poisson approximation).

We separately show the hit probability on the first cache (left plot)
and the hit probability of the entire cache network (right plot), observing
excellent agreement between analysis and simulation in all cases.
We note that LCP significantly outperforms LCE, as it
better exploits the aggregate storage capacity in the network
avoiding the simultaneous placement of the object in all caches.
Yet, LCD replication strategy performs even better, thanks to an improved  
filtering effect (LCD can be regarded as the dual of k-LRU for cache networks).

Then, we consider a very large topology comprising 1365 caches,
corresponding to a 4-ary regular tree with 6 levels. Such topology
is extremely expensive (if not impossible) to simulate,
whereas the model can predict its behavior at the same computation cost
of previous chain topology. Fig. ~\ref{fig:figura5}
reports the total hit probability achieved in this large network, 
for two traffic scenarios (analytical results only).

\tgifeps{12}{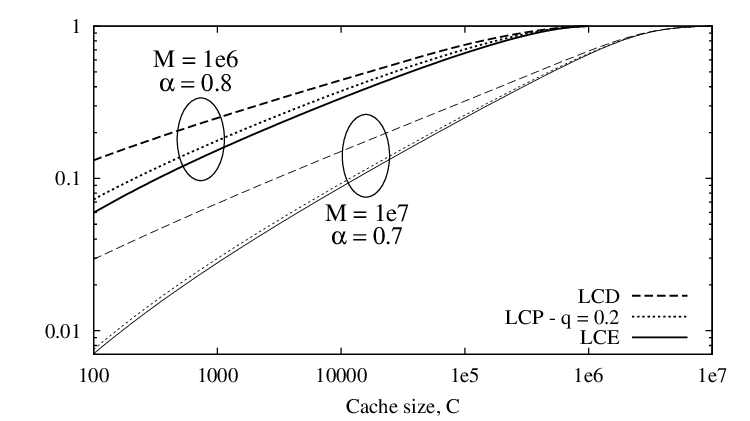}{Hit probability vs cache size, for various
replication strategies, in the case of a tree topology with 1365 caches, for two
traffic scenarios}

We again observe the huge gain of LCD with respect to LCE,
whereas the benefits of LCP are not very 
significant, especially with $\alpha = 0.7$.

\begin{figure}[tp]
\begin{minipage}[t]{0.3\linewidth}
\centering
\includegraphics[width=5cm]{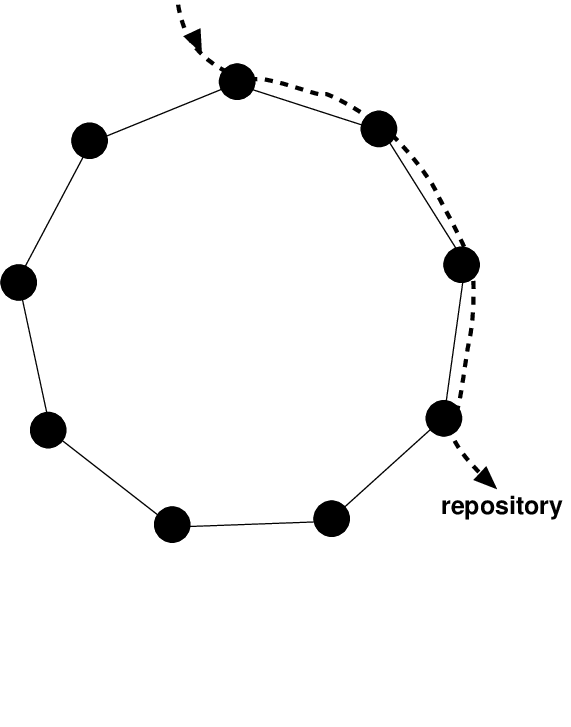}
\caption{Ring topology of 9 caches. The path followed
by requests entering one particular cache is shown as a
dashed line.}
\label{fig:ring}
\end{minipage}
\hspace{0.1\linewidth}
\begin{minipage}[t]{0.6\linewidth}
\centering
\includegraphics[width=10cm]{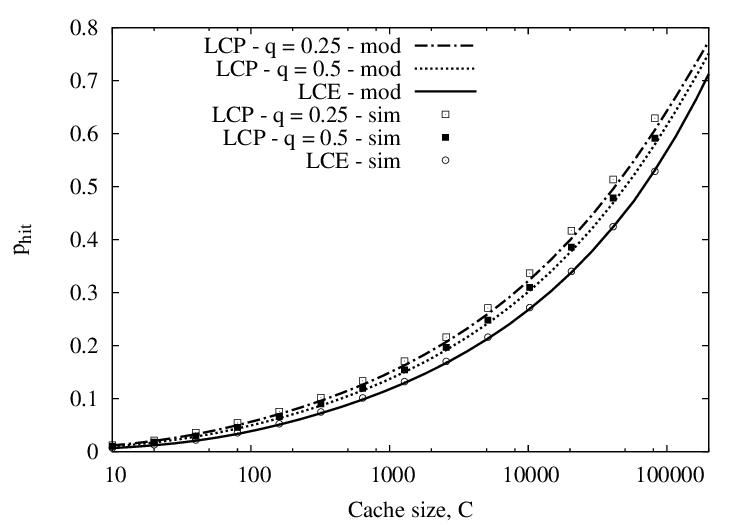}
\caption{Performance of LCE and LCP (with $q = 0.5$ or $q = 0.25$) in the ring topology. 
Comparison between analysis and simulation.}
\label{fig:figura7bis}
\end{minipage}
\end{figure}

At last, we consider an example of mesh network
comprising 9 caches arranged on a ring topology.
Requests  can enter the ring at any point, i.e., any of the caches 
along the ring acts as an ingress cache. Requests are forwarded
clockwise along the ring. However, requests that have traversed 4 caches 
without hitting the content are redirected to a remote, common 
repository storing all contents. Fig. \ref{fig:ring}
shows the path followed by the requests arriving externally
at one particular cache (requests entering the network
at the other caches are treated in a similar way).
The total external traffic of incoming requests is uniformly 
distributed over the 9 caches.

Fig. \ref{fig:figura7bis} compares the performance of LCE and LCP (with either $q = 0.5$ or $q = 0.25$)
in the considered mesh network, showing the global hit probability
achieved by the caching system. Here we have chosen the usual 
setting of $M=10^6$ and $\alpha=0.8$.
We have not considered in this scenario the LCD replication strategy, which
is primarily meant for hierarchical (tree-like) caching systems
and whose performance on general networks with cyclic topology are 
typically worse than LCP~\cite{rossi-icn14}.

Observe that also in the more challenging case of cache networks including cycles, 
the application of our model leads to pretty accurate 
predictions of the hit-probability. We wish to recall that networks 
which do not have feed-forward topology cannot be analyzed 
with existing techniques, such as that proposed in \cite{nain-CN}.

%% file: related.tex
\section{Related work}\label{sec:related}

The literature on caching systems is vast, so we limit ourselves to 
mentioning the papers more closely related to our work, mainly with a modeling flavour.
The first attempts to characterize the performance of simple caching systems date
back to the early 70's ~\cite{King,Gelenbe}. In \cite{King} authors have shown that 
the computational complexity of an exact model of a single LRU or FIFO cache 
grows exponentially with both the cache size $C$ and the catalogue size $M$.
In \cite{Gelenbe} it was proven that FIFO and RANDOM replacement policies
achieve exactly the same hit probability under IRM traffic.
Given that an exact characterization of most caching policies is prohibitive, 
approximated methodologies for the analysis of these systems have been 
proposed over the years \cite{Dan1990,Che}. The work~\cite{Dan1990} proposes 
an approximate technique with complexity $O(CM)$ for the estimation of the hit
probability in a LRU cache under IRM. The above technique can be extended also to FIFO 
caches, although in this case the asymptotic complexity cannot be precisely
determined due to the iterative nature of the model solution.  
A different approximation for LRU caches under IRM was originally proposed by \cite{Che}. 
This approximation constitutes the starting point of our work and it 
is explained in detail in Sec.~\ref{sect:che-approx}. 

Another thread of works by Jelenkovi\'{c}~\cite{Jele1,Jele2,Jele3,Jele4} has focused on the 
asymptotic characterization of the hit probability in LRU caches when the 
catalog size and the cache size jointly scale to infinite. In particular, \cite{Jele1} provides a 
closed form expression for the asymptotic hit probability in a large LRU cache under IRM
traffic with Zipf's exponent $\alpha>1$. Later works~\cite{Jele2,Jele3} have shown that LRU, 
in the asymptotic regime, exhibits an insensitivity property to traffic temporal locality. 
At last~\cite{Jele4} established the precise conditions on the scaling of parameters 
under which the insensitivity property holds. 
More recently, in~\cite{jel08}, the same author proposed the \lq\lq persistent-access-caching'' (PAC) 
scheme, showing that it provides nearly-optimal asymptotic performance under IRM with Zipf's exponent $\alpha>1$. 
We emphasize that the idea behind the PAC scheme shares some similarities with the $k$-LRU scheme
proposed in this work: under both schemes an insertion policy is added
to LRU to prevent unpopular contents from entering the cache. However, the configuration of 
PAC is harder as it requires setting several parameters. The $k$-LRU scheme, instead, is 
simpler and self-adapting.
{Other generalizations/extensions of LRU known as LRU-2Q,  LRU-$k$ and  LRFU have been proposed in \cite{LRU-2Q}, \cite{LRUK} and \cite{LRFU} respectively. LRU-2Q    is essentially  equivalent to  $k$-LRU, in  the case of $k=2$. 
Both LRU-$k$ and LRFU, instead, subsume either LRU or LFU by making the choice of the content to be evicted
dependent by the pattern of last $k$ observed content-requests.
$k$-LRU is somehow complementary to both LRU-$k$ and LRFU, since it enhances only the insertion policy 
of the classical LRU, by restricting access to the cache only to those contents which are sufficiently popular, 
while preserving the simplicity of LRU eviction.}

In the last few years cache systems have attracted renewed interest 
in the context of ICN. In \cite{pavlou-ccn} a Markovian approach has been proposed to approximate
the hit probability in LRU caches under IRM. The proposed method, however, is based on 
Markovian assumptions and cannot be easily extended to non-IRM traffic. 
In \cite{Carofiglio2011} the approach of~\cite{Jele1} has been extended 
to analyze the chunkization effect on cache performance in an ICN context.
An asymptotic characterization (for large caches) of the hit probability achieved by the
RANDOM policy is provided in ~\cite{muscariellosig}. {The trade-off between recency
and frequency in LRFU has been studied in \cite{innetworkingcaching}.}

The work \cite{Roberts1} provides a theoretical justification to Che's approximation for LRU,
and introduces a first attempt to apply Che's approach to non-LRU caches, 
considering the RANDOM policy under IRM. We emphasize that the approach 
proposed in \cite{Roberts1} to analyse RANDOM differs substantially from 
ours, being significantly more complex and hardly extendible to non-IRM traffic. 
At last, we wish to mention that Che's approximation for LRU has been very recently~\cite{amiciburini}  
extended to non-IRM traffic in special cases, adopting a dual approach with respect to ours.

With respect to all of the above-mentioned works, the goal of our paper is different,
in that here we show that the decoupling principle underlying Che's approximation
is much more general and flexible than what originally thought, 
and can be successfully applied to a broad set of caching policies 
under different traffic conditions, within a unified framework. 

For what concerns cache networks, we mention \cite{kurose2013,kurose2010,muscariellosig}. 
The work \cite{kurose2013} explores ergodicity conditions for arbitrary (mesh) 
networks. The models in \cite{kurose2010} and \cite{muscariellosig} rely on the independence
assumption among caches, assuming that requests arriving at each cache satisfy the IRM assumptions.
In contrast, we propose a methodology to capture the existing correlation 
among the states of neighboring caches, in a computationally efficient manner, 
considerably improving the accuracy of analytical predictions.
Our approach also permits analyzing cache networks adopting tightly coordinated 
replication strategies such as leave-copy-down.
We remark that cache networks implementing LCD have been previously considered in~\cite{lao06} for the special case
of tandem topologies. Our methodology provides a significantly simpler and 
higher scalable alternative to 
the approach devised in~\cite{lao06}, by capturing in a simple yet
effective way existing correlations between caches' states,
while reducing the number of parameters that must be estimated 
through fixed-point procedure.

Finally an alternative approach to ours has been recently 
proposed in~\cite{nain-CN},\cite{choungmofofackthesis},\cite{Fofack:value2} for cache networks 
with feed-forward topology, implementing TTL-based eviction policies. 
Their approach, which can be used to analyse the performance of LRU, RANDOM and FIFO under the Che approximation,  
essentially consists in characterizing the inter-request process arriving at non-ingress caches 
through a two steps procedure: i) the miss stream of (ingress) caches is exactly characterized  
as a renewal  process with given distribution; ii) by exploiting known results on the superposition of 
independent renewal processes, the exact inter-request time distribution at 
non-ingress caches is obtained.  Observe, however,  that the request  process at  non-ingress caches are, 
in general, non-renewal (since the superposition of independent renewal processes is not guaranteed to be renewal).
 Thus,  while the procedure proposed in~\cite{nain-CN}  is exact for network of TTL caches with linear topology, 
it can be  applied to  network of caches with tree-structure  only by approximating 
the  request processes  at non-ingress caches with renewal processes.
Recently a nice refinement of the approach followed by ~\cite{nain-CN} has been proposed in \cite{Berger},
where it  has shown  that the miss stream of TTL-based caches is a Markovian arrival process (MAP), 
provided that the request process is MAP. In light of the fact that the superposition of independent MAPs is also a MAP, 
\cite{Berger} has derived an exact analytical method for the analysis of feed-forward 
networks of TTL caches under MAP traffic.  

Although the approach in~\cite{nain-CN} and \cite{Berger} is very elegant, and 
can be potentially extended to renewal traffic, it suffers from the following two limitations:
i) it becomes computationally very intensive when applied to large networks;
ii) it can be hardly generalized to general mesh networks (non feed-forward). 
Our approach is somehow complementary to the one followed by~\cite{nain-CN} and \cite{Berger} since, 
while it applies only to IRM traffic, it is much more scalable and 
readily applicable to networks with general topology. 


%% file: appendix.tex
\section{Proof of q-LRU $\to$ LFU: IRM case}\label{app:qlru}
We first prove that $\lim_{q\to 0} T_C=+\infty$. Consider function $f(T_C,q)\triangleq \sum_{m}p_{\text{in}}(m)$, 
From \eqref{eq:C}, $f(T_C,q)\equiv C$.
Recalling \eqref{eq:q-LRU}, we have:
\[
f(T_C,q)=\sum_m\dfrac{q(1- e^{-\lambda_m T_C})}{e^{-\lambda_m T_C}+q(1- e^{-\lambda_m T_C})}
\]
where  previous sum extends over all contents in the catalog (which is assumed to be of finite size $M$).
Deriving the above formula, we obtain:
\begin{equation}
\label{eq:df_q}
f_q\triangleq\dfrac{\partial f}{\partial q} = 
\sum_m \dfrac{(1- x)(x+q(1- x))-q(1- x)^2}{(x+q(1- x))^2}\bigg|_{x=e^{-\lambda_m T_C}}
= \sum_m \dfrac{(1- x)x}{(x+q(1- x))^2}\bigg|_{x=e^{-\lambda_m T_C}}>0
\end{equation}
and
\begin{multline}
f_{T_C} \triangleq \dfrac{\partial f}{\partial T_C} = \dfrac{\partial f}{\partial x}\bigg|_{x=e^{-\lambda_m T_C}}\dfrac{\partial e^{-\lambda_m T_C}}{\partial T_C}
=\sum_m\partial\left( \dfrac{q(1-x)}{x+q(1-x)}\right)/\partial x\bigg|_{x=e^{-\lambda_m T_C}}
(-\lambda_m e^{-\lambda_m T_C}) = \\
=\sum_m \left.\dfrac{-q(x+q(1-x))-q(1-x)(1-q)}{(x+q(1-x))^2} \right|_{x=e^{-\lambda_m T_C}}
(-\lambda_m e^{-\lambda_m T_C}) = \\
=\sum_m \dfrac{q\lambda_m e^{-\lambda_m T_C}}{(e^{-\lambda_m T_C}+q(1-e^{-\lambda_m T_C}))^2}>0
\end{multline}
By the implicit function theorem, we have 
\[
 \dfrac{\partial T_C}{\partial q}=\dfrac{-f_q(T_C,q)}{f_{T_C}(T_C,q)}<0
\]

We can conclude that $T_C$ is a decreasing function with respect to $q$, thus we have 
that the limit $\lim_{q\to 0} T_C$ exists and equals $\sup_q T_C$. 
{
We prove now that this limit 
is indeed equal to infinity.
We define \mbox{$T_{C,\text{sup}} \triangleq  \text{sup}_q T_C(q)=\lim_{q\to 0} T_C$}, and we suppose by contradiction that 
this is a finite quantity. In this case, we would have
{
\[
 \lim_{q\to 0} f(T_C,q) = \lim_{q\to 0}\sum_m p_{\text{in}}(m)=\lim_{q\to 0}\sum_{m}\dfrac{q(1- e^{-\lambda_m T_C})}{e^{-\lambda_m T_C}+q(1- e^{-\lambda_m T_C})}=0,
\]
}
in contrast with the fact that the previous sum is equal to $C$, by definition. Thus, { $T_{C,\text{sup}}\triangleq \lim_{q\to 0}T_C(q)=+\infty$.}
We prove now that $T_C(q)$ asymptotically behaves  as $c \log \frac{1}{q}$ for some $c>0$ as $q\to 0$.
We can write:
\begin{eqnarray}
\label{eq:TC_log}
\begin{array}{l}
  \lim_{q\to 0}\sum_m p_{\text{in}}(m)=\\
   \lim_{q\to 0} \sum_m \dfrac{q(1-e^{-\lambda_mT_C}))}{e^{-\lambda_mT_C}+q(1-e^{-\lambda_mT_C})}=\\
\lim_{q\to 0} \sum_m \dfrac{q+o(q)}{e^{-\lambda_mT_C}+q+o(q)}=\\
\lim_{q\to 0} \sum_m \dfrac{1+o(1)}{1+e^{-\lambda_mT_C}/q+o(1)}=\\
\lim_{q\to 0} \sum_m \dfrac{1+o(1)}{1+e^{-(\lambda_mT_C-\log (1/q))}+o(1)}
\end{array}
\end{eqnarray}
We note that, if $\frac{T_C}{\log(1/q))}$ becomes arbitrarily large  as $q\to 0$, every term in \eqref{eq:TC_log} 
tends to 1, and the sum would be equal to the number of contents, whereas we know that it has to 
be equal to $C$. If, on the other hand, 
$\frac{T_C}{\log(1/q))}$ becomes arbitrarily small as $q\to 0$,  every term in the sum in \eqref{eq:TC_log} would tend 
to 0. We can thus conclude that  $\frac{T_C}{\log(1/q)}$  is bounded away from both 0 and infinite.

Thus assuming for the moment that $\lim_{q\to 0} \frac{T_C}{\log(1/q))}$ exists, it must   necessarily be equal to $c>0$.
Now, by setting $\lambda^*=1/c$ we have:  
\[
\lim_{q \to 0} p_{\text{in}}(m)=\lim_{q\to 0}\dfrac{1+o(1)}{q^{\frac{\lambda_m}{\lambda^{*}}-1}+1+o(1)}=
\left\{\begin{array}{rl}1 & \text{if } \lambda_m\geq \lambda^{*}\\ 0 & \text{if } \lambda_m<\lambda^{*} \end{array}\right.
\]
{{
Note that previous argument still holds when $\lim_{q\to 0} \frac{T_C}{\log(1/q))}$  does not exist, 
provided that the following  condition is met: i) no $\lambda_m$  can be found,
with $\lambda^*  < \lambda_m \le \Lambda^*$, such that
 $0<\liminf_{q\to 0}  \frac{T_C}{\log(1/q))} =\frac{1}{\Lambda^*} < \limsup_{q\to 0}  \frac{T_C}{\log(1/q))}=\frac{1}{\lambda_m}<\infty$.
} 
}

{ At last we show, by contradiction, that either   $\lim_{q\to 0} \frac{T_C}{\log(1/q))}$ exists or 
condition i) above is met.  Indeed, assume that there is  an $m$ such that   $\lambda^*  \le \lambda_m < \Lambda^*$.
Then, denoting with $\Uno_{\{A \} }$ the indicator function associated to the event $\{A\}$,
 by construction it must be both
 $\sum_{m} \Uno_{ \{ \lambda_m\ge \lambda^*\} }=C$  and $\sum_{m}\Uno_{ \{ \lambda_m \ge \Lambda^* \} }=C$.
Thus:   $\sum_{m} \Uno_{ \{ \lambda_m\ge \lambda^*\} }= \sum_{m}\Uno_{ \{ \lambda_m \ge \Lambda^* \} }$, which is in contradiction with the assumption.}

\section{Proof of q-LRU $\to$ LFU: general case}\label{app:qlru_general}
To simplify the proof we assume the support of the inter-request time pdf to be simply connected. 
As consequence, $F(m,y)$  ($\hat{F}(m,y)$) is a  strictly increasing function with respect to variable $x$ ($y$)  on its relevant range, i.e, for any $x$ such that $0<F(m,x)<1$ ( $\forall\ y$ s.t. $0<\hat{F}(m,y)<1$). First we consider to the case in which $F(m,x)$  ($\hat{F}(m,y)$) has an infinite support for any $m$.
In this case we first prove that $\lim_{q\to 0} T_C=+\infty$. Consider function $f(T_C,q)\triangleq \sum_{m}p_{\text{in}}(m)$.
From \eqref{eq:C}, $f(T_C,q)\equiv C$.
Recalling \eqref{eq:qpin-LRU_implicit-gen} and \eqref{eq:qphit-LRU_implicit-gen}, we have:
\[
 p_{\text{hit}}(m)=\dfrac{q F(m,T_C)}{1-F(m,T_C)(1-q)}
\]
and
\begin{multline}
p_{\text{in}}(m) = \hat{F}(m,T_C)[p_{\text{hit}}(m)+q(1-p_{\text{hit}}(m))] = \\
\hat{F}(m,T_C)\left[\dfrac{q F(m,T_C)}{1-F(m,T_C)(1-q)} + q \left(1-\dfrac{q F(m,T_C)}{1-F(m,T_C)(1-q)} \right)\right] = 
\hat{F}(m,T_C)\dfrac{q}{1-F(m,T_C)(1-q)}
\label{eq:qpin-LRU_expl-gen}
\end{multline}

Thus, 
$$f(T_C,q)=\sum_m \hat{F}(m,T_C)\dfrac{q}{1-F(m,T_C)(1-q)}.$$
Deriving this formula, we obtain:
$$ f_q \triangleq \dfrac{\partial f}{\partial q}=\sum_m  \hat{F}(m,T_C) \dfrac{1-F(m,T_C)(1-q)-qF(m,T_C)}{\left[1-F(m,T_C)(1-q)\right]^2} = \sum_m  \hat{F}(m,T_C) \dfrac{1-F(m,T_C)}{\left[1-F(m,T_C)(1-q)\right]^2}>0$$
and
\begin{multline*}
 f_{T_C} \triangleq \dfrac{\partial f}{\partial T_C} = 
 \sum_m \dfrac{\partial \hat{F}(m,T_C) }{\partial T_C} \left[\dfrac{q}{1-F(m,T_C){(1-q)}} \right]
+ \hat{F}(m,T_C)\dfrac{\partial }{\partial T_C}\dfrac{q}{1-F(m,T_C)(1-q)} = \\
= \sum_m \dfrac{\partial \hat{F}(m,T_C) }{\partial T_C} \left[\dfrac{q}{1-F(m,T_C){(1-q)}} \right]
+ \hat{F}(m,T_C)\dfrac{q(1-q)}{\left[1-F(m,T_C)(1-q)\right]^2} \dfrac{\partial {F}(m,T_C) }{\partial T_C}>0,
\end{multline*}
since both $\hat{F}(m,T_C)$ and $F(m,T_C)$ are increasing with $T_C$.

By the implicit function theorem, we have 
\[
 \dfrac{\partial T_C}{\partial q}=\dfrac{-f_q(T_C,q)}{f_{T_C}(T_C,q)}<0
\]
We can conclude that $T_C$ is a decreasing function with respect to $q$, thus we have that 
the limit $\lim_{q\to 0} T_C$ exists and equals $\sup_q T_C$ . We prove now that this limit 
is indeed equal to infinity.
We define $T_{C,\text{sup}} \triangleq \text{sup}_q T_C(q)= \lim_{q\to 0} T_C$, and we suppose, by contradiction, that this is a finite quantity.
In this case, we would have
\[
{\lim_{q\to 0} f(T_C,q)}= \lim_{q\to 0}\sum_m p_{\text{in}}(m)=\lim_{q\to 0}\sum_{m}\hat{F}(m,T_C)\dfrac{q}{1-F(m,T_C)(1-q)}=0,
\]
in contrast with the fact that the previous sum is equal to $C$, by definition. 
{
Thus, since  $\lim_{q\to 0}T_C(q)=+\infty$  we have:
\begin{equation}
 \lim_{q\to 0}p_{\text{in}}(m)=\lim_{q\to 0}\sum_{m}\hat{F}(m,T_C)\dfrac{1}{[1-F(m,T_C)(1-q)]/q},
\label{eq:pLRU-LFU}
\end{equation}
now, observe that  if $1-F(m,T_C)=o(q)$, the previous limit becomes equal to $1$, 
whereas, if $1-F(m,T_C)=\omega(q)$, the limit is equal to $0$.

\tgifeps{8}{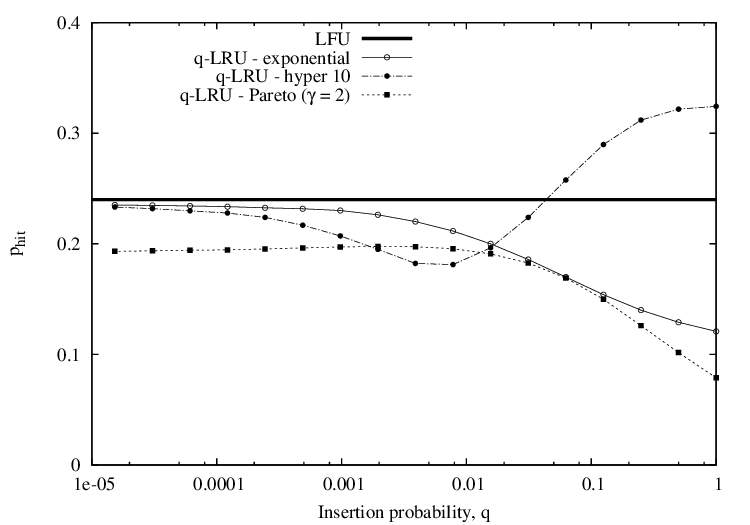}{Hit probability vs insertion probability of q-LRU, 
for different inter-request time distributions, fixed cache size equal to 10,000,
$\alpha=0.7$.}

Then, with similar arguments as for the exponential case, {under our assumptions (i.e.,  
the fact that we assume $\lim_{t\to \infty} \frac{1-F(m_1,t)}{1-F(m_2,t)}=\infty$  whenever $\lambda_{m_1}<\lambda_{m_2}$)}
we can easily show that there necessarily exists some $m_0$  such that 
$1-F(m,T_C)=o(q)$  for $m<m_0$ and  $1-F(m,T_C)=\omega(q)$  for $m>m_0$. Indeed  observe that, by hypothesis, 
the asymptotic behavior of  $1-F(m,T_C)$ as   $T_C \to \infty$  depends  on $m$ 
(i.e., on arrival rates $\lambda_m$'s, which are assumed to be different for different $m$).

Fig. \ref{fig:extra} provides a numerical confirmation of our
theoretical predictions (see also Remark after Theorem \ref{theo:qlrugen}),
plotting the hit probability as function of the insertion probability of q-LRU
under different inter-request time distributions: exponential, hyper-10, 
Pareto (with exponent $\gamma = 2$). This experiment suggests that both
the exponential and hyper-10 curves approach LFU as $q \rightarrow 0$, while
the curve corresponding to the Pareto case tends to a different limit.

The case in which $F(m,T_C)$ has a bounded support  for some $m$ can be  treated similarly. 
However if  the number of contents with finite support exceeds $C$,  $T_C$ does not tend anymore to $\infty$. 
Observe indeed that from \eqref{eq:pLRU-LFU}
we can deduce that every content  whose inter-request time has a maximum value, which is   smaller that $T_{C,\text{sup}}$,
will be necessarily found in the cache with 
a probability tending to 1 when $q\to 0$, while every other content will be found with a probability tending to 0. Thus, 
since by construction we have  $\sum_m P_{\text{in}}(m)=C$,
   only $C$ contents can have maximum inter-request time smaller than 
$T_{C,\text{sup}}$. This concludes the proof.}

%% file: app_2LRUp_in.tex
\section{Exact calculation of $p_{\text{in} (m)}$ for 2-LRU}\label{app:2LRUp_in}

For simplicity in this appendix, whenever not strictly necessary, we omit the dependency of variables on $m$.
We define as cycle the time interval between two visits at state ($1,1$) (\emph{i.e.}, the time interval between 
two requests for object $m$ that bring the system to state ($1,1$)).  
Observe that, by construction, the cycles are i.i.d. 
We consider a generic cycle starting at time $t=0$ (thus by construction a request for $m$ arrives at time $t=0$).
Let  $R_1$  be the time  of the first request for object $m$ after $t=0$. We have the following possibilities:
\begin{itemize}
 \item $R_1\leq T_C^1$:  at time $R_1^-$ the system  is still in  state ($1,1$), and consequently  
$\mathbb{E}[T_{\text{cycle }} \mid{R_1<T_C^1} ]= \mathbb{E}[R_1 \mid{R_1<T_C^1} ]$.

\item $T_C^1<R_1\leq  T_C^2$: in this case, at time $t=T_C^1$ the system enters  state ($0,1$), where it is found at $R_1^-$;
thus the request at $R_1$ brings the system again in state ($1,1$). In this case
${\mathbb{E}[T_{\text{cycle }}\mid{ T_C^1<R_1\leq  T_C^2}]}= \mathbb{E}[R_1 \mid{T_C^1<R_1\leq  T_C^2} ]$.

\item $R_1>T_C^2$: in this last case  the analysis is  more complicated. At time $T_C^1$ the system goes to state ($0,1$), 
and at time $T_C^2$  it enters  state ($0,0$). At time $t=R_1$,  for effect of 
the arrival of the  first  request, the system enters ($1,0$). 
Now, if the  following  request arrives  before  $R_1+T_C^1$, 
the system goes  back to state ($1,1$) and the cycle terminates; otherwise, the system  at time  $R_1+T_C^1$ enters 
again  state ($0,0$) and the following request brings it again to state ($1,0$).  
The cycle ends upon the  arrival of the first request for  object $m$ 
that follows the previous one by less than $T_C^1$. Figure \ref{fig:fig_appendice} better illustrates this situation.
\end{itemize}

\tgifeps{9}{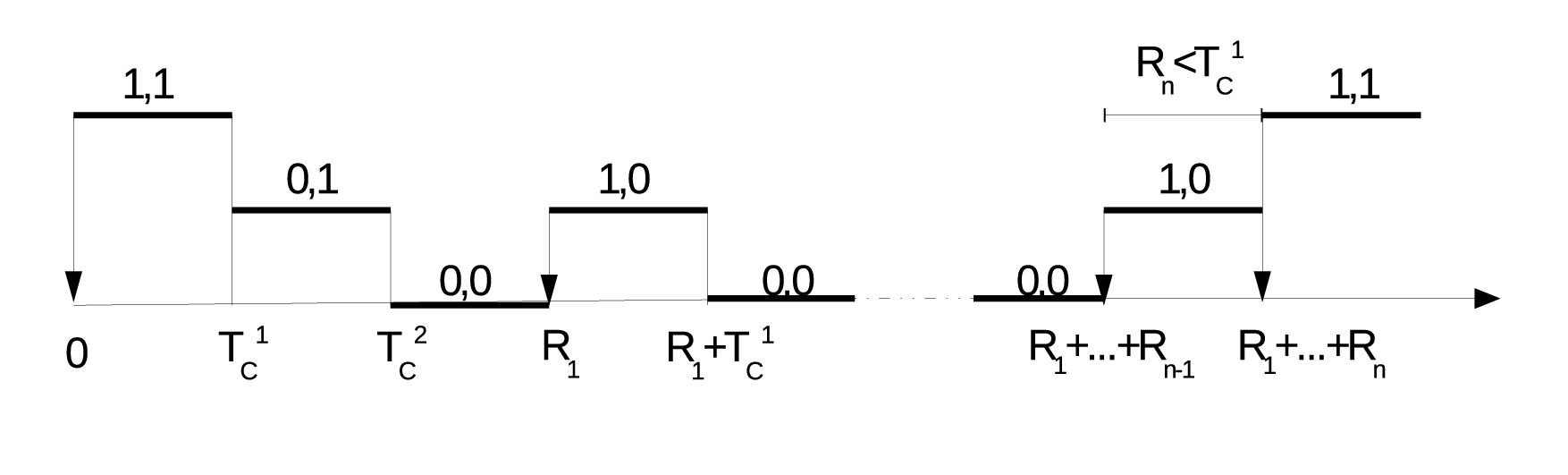}{Illustration of the cycle analysis used for deriving \eqref{eq:Tcycle}.Vertical arrows represent incoming requests for the content.}

{
Thus, if we denote by $R_i$ the $i$-th inter-request time, and with $n\sim \text{Geom}(p_1)$, 
$p_1=\mathbb{P}(R\leq T_C^1)=1-e^{-\lambda_m T_C^1}$ and $p_2=\mathbb{P}(R\leq T_C^2)=1-e^{-\lambda_m T_C^2}$, 
we can write in this case:
\begin{align*}
\mathbb{E}[T_{\text{cycle}}\mid{R_1 > T_C^2}] &= \mathbb{E}[R_1\mid{R_1>T_C^2}]+\mathbb{E}[R_n\mid{R_{n}\leq T_C^1}] 
+\mathbb{E}[\sum_{i=0}^{n-1} R_i\mid{R_i>T_C^1} ]\\
&=\mathbb{E}[R_1 \mid{R_1 > T_C^2}]+\mathbb{E}[R_n\mid{R_{n}\leq T_C^1}]+\mathbb{E}[n]\mathbb{E}[ R_i\mid{R_i>T_C^1} ]  \\
&= \mathbb{E}[R_1 \mid{R_1 > T_C^2}]+\dfrac{\mathbb{E}[R_n, R_{n}\leq T_C^1]}{\mathbb{P}(R_{n}\leq T_C^1)} +  \dfrac{1-p_1}{p_1}\dfrac{\mathbb{E}[ R_i,R_i>T_C^1 ]}{\mathbb{P}(R_{i}> T_C^1)}  \\
&=\mathbb{E}[R_1 \mid{R_1 > T_C^2}]+\dfrac{\mathbb{E}[R_n, R_{n}\leq T_C^1]}{p_1}  + \dfrac{1-p_1}{p_1}\dfrac{\mathbb{E}[ R_i,R_i>T_C^1 ]}{1-p_1}  \\
&=\mathbb{E}[R\mid{R > T_C^2}]+\dfrac{\mathbb{E}[R]}{p_1}
\end{align*}
}
{
Considering also the other cases, we have:
\begin{align}
\label{eq:Tcycle}
\mathbb{E}[T_{\text{cycle}}] &= \mathbb{E}[R_1\mid{R_1 < T_C^2}]\mathbb{P}(R_1\leq T_C^2)+ 
\left(\mathbb{E}[R_1\mid{R_1> T_C^2}]+\dfrac{\mathbb{E}[R]}{p_1}\right)\mathbb{P}(R_1 > T_C^2) \nonumber\\
&= \mathbb{E}[R]+\dfrac{\mathbb{E}[R]}{p_1}(1-p_2)
\end{align}
Turning our attention to $\mathbb{E}[T_{\text{ON}}]$, which is the average time within a cycle 
during which content $m$ is stored in the second (physical) cache, we have:
\begin{equation}
 \mathbb{E}[T_{\text{ON}}] = \mathbb{E}[ \min(R_1,T_C^2)]= \mathbb{E}[R_1\mid{R_1< T_C^2}]\mathbb{P}(R_1< T_C^2) + T_C^2 \,\mathbb{P}(R_1\ge T_C^2) 
\end{equation}
At last we can obtain $p_{\text{in}}(m)$ as: 
\[
 p_{\text{in}}(m)= \dfrac{\mathbb{E}[T_{\text{ON}}(m)]}{\mathbb{E}[T_{\text{cycle}}(m)]}
\]
}
\section{Proof of k-LRU $\to$ LFU}\label{app:klru}
For simplicity we limit ourselves to the IRM traffic model. An analogous result can be derived
under {\em renewal} traffic along the same lines.
First we recall that  sequence $\{T_C^i\}_{i=1}^{k}$ is  increasing.
We prove that $T_C^*=\sup_{k\to \infty} T_C^k=+\infty$. Indeed, assume by contradiction that $T_C^*$ is finite. 
Now, a necessary condition for content $m$ to be 
in the cache at time $t$ is that a  request arrived at $ \tau_1 \in ( t- T_C^k, t]$; this  request  in turn must have necessarily generated a hit  either in cache $k$ or in cache $k-1$. As consequence, a previous request  must have arrived at
 $\tau_{2} \in ( \tau_1- T_C^k, \tau_1]$.  Iterating back we  generate a chain of $k$ 
 requests for object $m$ requests with inter-request time smaller than
$T_C^k$, which is necessary for object $m$ 
to be found in cache $k$ at time $t$.
The probability  of observing such a chain is bounded by $(1-e^{-\lambda_m T_C^*})^k$,  and
this probability goes to zero when $k\to \infty$, independently on $\lambda_m$, leading to a contradiction. 
Indeed recall that, by construction, $\sum p_{\text{in}} (m,k)=C$. 
Thus, we can conclude that $\lim_{k\to \infty} T_C=+\infty$.
Recalling the expression in \eqref{eq:nLRU}: 
\[
 p_{\text{in}} (m,i)=(1- e^{-\lambda_m T_{C}^i)}) [ p_{\text{in}}(m,i)+( p_{\text{in}}(m,i-1))(1-p_{\text{in}}(m,i))] 
\]
we can easily prove that: $p_{\text{in}} (m,i)$  is increasing with respect to $\lambda_m$  for any $i$ (by induction over $i$); 
ii)  \eqref{eq:nLRU}, for sufficiently large  $T_C^i$, is a contraction mapping  over $[\epsilon,1]$ for any $\epsilon>0$ .

Thus,  $\lim_{k\to \infty} p_{\text{in}} (m,k)$   exists  and   it is necessarily  the fixed point 
 $p_{\text{in}}^* (m)$ of  \eqref{eq:nLRU}.    The assertion immediately  follows, since  
 $p_{\text{in}}^* (m)\in \{0,1\}$.

The extension to the non-IRM case,under the assumption that  the support of the 
inter-request time distribution is unbounded, and that for any $m_1$ and $m_2$, with 
 $\lambda_{m_1}<\lambda_{m_2}$, $\lim_{t\to \infty} \frac{1-F(m_1,t)}{1-F(m_2,t)}>1$,
 follows the same lines.

